\documentclass[12pt]{article}
\usepackage[latin1]{inputenc}
\usepackage{cite}
\usepackage{amsmath}
\usepackage{amsfonts}
\usepackage{amssymb}
\usepackage{graphicx}
\usepackage{geometry}
\usepackage{amssymb,epsfig}
\usepackage{hyperref}

\makeatletter
\renewcommand\section{\@startsection {section}{1}{\z@}%
                                 {-3.5ex \@plus -1ex \@minus -.2ex}
                                   {2.3ex \@plus.2ex}%
                                   {\normalfont\large\bfseries}}
\renewcommand\subsection{\@startsection{subsection}{2}{\z@}%
                                   {-3.25ex\@plus -1ex \@minus -.2ex}%
                                     {1.5ex \@plus .2ex}%
                                     {\normalfont\bfseries}}
\renewcommand\subsubsection{\@startsection{subsubsection}{3}{\z@}%
                                   {-3.25ex\@plus -1ex \@minus -.2ex}%
                                     {1.5ex \@plus .2ex}%
                                     {\normalfont\itshape}}
\makeatother

\def\pplogo{\vbox{\kern-\headheight\kern -29pt
\halign{##&##\hfil\cr&{\ppnumber}\cr\rule{0pt}{2.5ex}&\ppdate\cr}}}
\makeatletter
\def\ps@firstpage{\ps@empty \def\@oddhead{\hss\pplogo}%
  \let\@evenhead\@oddhead 
}
\def\maketitle{\par
 \begingroup
 \def\thefootnote{\fnsymbol{footnote}}
 \def\@makefnmark{\hbox{$^{\@thefnmark}$\hss}}
 \if@twocolumn
 \twocolumn[\@maketitle]
 \else \newpage
 \global\@topnum\z@ \@maketitle \fi\thispagestyle{firstpage}\@thanks
 \endgroup
 \setcounter{footnote}{0}
 \let\maketitle\relax
 \let\@maketitle\relax
 \gdef\@thanks{}\gdef\@author{}\gdef\@title{}\let\thanks\relax}
\makeatother

\numberwithin{equation}{section}

\renewcommand{\th}{\theta}

\newcommand{\be}{\begin{equation}}
\newcommand{\bea}{\begin{eqnarray}}
\newcommand{\ee}{\end{equation}}
\newcommand{\eea}{\end{eqnarray}}

\begin{document}
 
\setcounter{page}0
\def\ppnumber{\vbox{\baselineskip14pt
}}
\def\ppdate{\footnotesize{}} \date{}

\author{Carlos Tamarit\\
[7mm]
{\normalsize  \it Perimeter Institute for Theoretical Physics}\\
{\normalsize \it Waterloo, ON, N2L 2Y5, Canada}\\
[3mm]
{\tt \footnotesize ctamarit at perimeterinstitute.ca}
}

\bigskip
\title{\bf Decoupling heavy sparticles in Effective SUSY scenarios: Unification, Higgs masses and tachyon bounds 
\vskip 0.5cm}
\maketitle

\begin{abstract} \normalsize
\noindent 
Using two-loop renormalization group equations implementing the decoupling of heavy scalars, Effective SUSY scenarios are studied in the limit in which there is a single low energy Higgs field. Gauge coupling unification is shown to hold with similar or better precision than in standard MSSM scenarios. $b$-$\tau$ unification is examined, and  Higgs masses are computed using the effective potential, including two-loop contributions from scalars. A 125 GeV Higgs is compatible with stops/sbottoms at around 300 GeV with non-universal boundary conditions at the scale of the heavy sparticles if some of the trilinear couplings at this scale take values of the order of 1-2 TeV; if more constrained boundary conditions inspired by msugra or gauge mediation are set at a higher scale, heavier colored sparticles are required in general. Finally, since the decoupled RG flow for third-generation scalar masses departs very significantly from the MSSM $\overline{\rm DR}$ one, tachyon bounds for light scalars are revisited 
and shown to be relaxed by up to a TeV or more.

\end{abstract}
\bigskip
\newpage
\tableofcontents
\newpage

\section{Introduction}

Given the bounds that the LHC experiments are inducing in constrained or partly degenerate parameter subspaces of the Minimal Supersymmetric Standard Model (MSSM) (see for example \cite{Buchmueller:2011sw,Kats:2011qh}), if Supersymmetry (SUSY)
is realized in Nature and moreover solves the Hierarchy problem, it may involve a non-universal/hierarchical spectrum of superpartners. Among possible natural SUSY scenarios, Effective SUSY  \cite{Dimopoulos:1995mi,Cohen:1996vb} is very compelling. These scenarios have light fermion superpartners and light third-generation scalars, which is enough to render the Higgs sector natural. First and second generation scalars are heavy and may reach up to 20 TeV without spoiling naturalness, which in turn helps to solve the flavor problem via decoupling.

The minimal Effective SUSY scenario solving the naturalness problem has as light third generation scalars only the left handed squark doublet and the right handed stop \cite{Brust:2011tb}. One may also have non-minimal scenarios, and perhaps the simplest to realize in models addressing SUSY breaking is that in which all the third generation scalars remain light. 

Given the large hierarchies in the soft masses in Effective SUSY models, it has been pointed out \cite{Tamarit:2012} that perturbation theory using the MSSM $\overline{\rm DR}$ RG equations will become problematic due to the appearance of large logarithms in the finite quantum corrections. This effect is particularly significant in the RG flow of the soft masses of the light third generation scalars. It is known that two-loop effects due to the heavy first and second generation scalars tend to drive the light soft masses to negative values \cite{ArkaniHamed:1997ab}. However, the heavy sparticles also give rise to very large and positive finite corrections. Not only is the computation of the latter problematic when the tree-level value of the masses is tachyonic, but also the large size of these corrections puts into question the reliability of perturbation theory and the precision of the resulting values for the physical masses.

All these problems stem from the use of an unphysical renormalization scheme such as $\overline{\rm DR}$, which does not implement the decoupling of heavy particles. A way out  relies on obtaining the RG equations for the effective theories in which the heavy scalars are decoupled and then matching these theories with the MSSM at the scale of the heavy sparticles. The two-loop RG equations for Effective SUSY scenarios with heavy scalars decoupled and a single  Higgs field at low energy have already been obtained in ref.~\cite{Tamarit:2012}. 

The aim of this paper is to use the results of ref.~\cite{Tamarit:2012} to perform a two-loop analysis of some basic properties of Effective SUSY scenarios with a Standard Model-like Higgs. Flavour mixing effects and complex phases are not taken into account. The topics addressed are gauge and Yukawa coupling unification, Higgs masses, fine-tuning estimates and bounds on soft masses obtained by demanding that there are no colored vacua. 

Regarding unification, it is known that the MSSM $\overline{\rm DR}$ couplings at two-loops unify to great precision, of around 1-3\%, requiring negative threshold corrections at the GUT scale for the strong coupling \cite{Pierce:1996zz}. In Split SUSY scenarios --in which all scalars of the MSSM are made heavy-- a similar analysis using decoupled 2-loop RG flows shows that unification is improved over a range of scales for the heavy sparticles, requiring smaller threshold corrections \cite{Giudice:2004tc}. It will be shown that Effective SUSY scenarios not only do not spoil the standard MSSM unification but may improve upon it: unification is typically achieved with a 1-2\% accuracy, also requiring negative threshold corrections for $g_3$ at the GUT scale. In the case of $b$-$\tau$ unification, necessary for some SU(5) GUT theories, it will be shown that it is hard to obtain in minimal Effective SUSY scenarios, but possible in nonminimal ones when nonzero sbottom mixing angles are present.

Given the current strengthening evidence for a Higgs boson with mass around 125-126 GeV \cite{ATLAS:2012ae,Chatrchyan:2012tx}, this paper presents computations of Higgs masses in Effective SUSY scenarios with a single Higgs field, relying on the two-loop RG equations as well as the one-loop effective potential supplemented by the two-loop contributions due to scalars. The results show that a 125 GeV Higgs is possible in these scenarios, even with stops/sbottoms around 300 GeV, if the  boundary conditions at the scale of the heavy particles are non-universal and trilinear couplings at this scale get large, of the order of 1-2 TeV. If boundary conditions inspired by msugra or gauge mediation are imposed at higher scales, heavier stops are typically required. These results are qualitatively similar to those obtained in the MSSM without imposing 
an inverted hierarchy in the squark sector \cite{Carena:2011aa,Heinemeyer:2011aa,Arbey:2011ab,Hall:2011aa,Draper:2011aa}; the present analysis also contemplates the possibility of large $a$ terms in the down sector, which may also give rise to a 125 GeV Higgs with light stops/sbottoms even with small or moderate values of $a_u$; this signals that quantum corrections coming from the down sector can become relevant. Also, when using boundary conditions inspired by gauge mediation for the third generation scalars and gauginos (introducing separate scales for scalar and fermion masses), with zero trilinear couplings at the susy breaking scale, a 125 GeV is possible with stops/sbottoms around 2-3 TeV, much lighter than the 10 TeV required in traditional gauge mediated scenarios with universal scalar masses for the three generations \cite{Draper:2011aa}. Fine-tuning estimates are also provided; it is shown that models with large a-terms and light stop/sbottoms have a fine-tuning of the order of one part in 200-300,
 if the measure of eq.~\eqref{eq:FTmeasure} is used.

Concerning colored vacua, as was mentioned earlier two-loop RG effects due to the heavy sparticles are known to drive the third generation soft masses towards tachyonic values. In the case of high-scale SUSY breaking, this usually produces tachyonic stops, while in the case of low scale SUSY breaking scenarios, such as models involving gauge mediation, it may give rise to tachyonic sleptons. Demanding no tachyons produces lower bounds for the third generation soft masses. As has been argued, the RG flow of the latter changes significantly after implementing decoupling, so that the bounds have to be revisited.  It will be shown that taking into account the decoupling of heavy particles the bounds are relaxed by up to 900 GeV in models of high scale SUSY breaking and 1.5 TeV in the case of low-scale breaking.

The rest of the paper is organized as follows. Section \ref{sec:Lagrangians} introduces the Lagrangians of the effective theories with the heavy sparticles decoupled, following ref.~\cite{Tamarit:2012}. Gauge coupling and $b$-$\tau$  unification are analyzed in \S\S~\ref{sec:Unification} and \ref{sec:btauunif}, respectively. \S \ref{sec:Higgs} is dedicated to Higgs masses,  while tachyon bounds are treated in \S \ref{sec:Tachyons}. Three appendices are included. \S \ref{app:highscale} explains the boundary conditions for the effective theories at the scale of the heavy scalars by matching the low energy couplings with those of the MSSM in a Higgs decoupling limit, taking into account threshold corrections due to the transition between $\overline{\rm MS}$ and $\overline{\rm DR}$ schemes. \S \ref{app:lowscale} summarizes the one-loop threshold corrections used at the scale of the top mass and at the lower Effective SUSY threshold at which the Standard Model couplings are matched with those of the Effective 
SUSY theories. Finally, \S \ref{app:VCW} summarizes the computation of the effective potential including two-loop effects due to scalars.

\section{Low energy Lagrangians\label{sec:Lagrangians}}

This section introduces the field content and Lagrangians of the minimal and nonminimal Effective SUSY scenarios described before. Throughout this paper, fermion fields are denoted with lower case letters, and scalars with upper case ones.
\subsection{Minimal Effective SUSY\label{subsec:Lagrangians:MinEffSUSY}}

The low energy theory contains the Standard Model fields including the Higgs, gauginos, two higgsinos, the third generation left-handed squark doublet and the right handed stop of the MSSM. The field content is summarized in the next table.

\begin{center}
\begin{tabular}{c|c|c|c}
 \   & SU(3) & SU(2) & U(1) \\
\hline  
$q_i,i=1\dots 3$ & $\square$ & $\square$ & 1/6 \\
\hline
$u^c_{i},i=1\dots 3$ & $\overline\square$ & $\mathbb{I}$ & -2/3 \\
\hline
$d^c_{i},i=1\dots 3$ & $\overline\square$ & $\mathbb{I}$ & 1/3 \\
\hline
$l_{i},i=1\dots 3$ & $\mathbb{I}$ & $\square$ & -1/2 \\
\hline
$e^c_{i},i=1\dots 3$ & $\mathbb{I}$ & $\mathbb{I}$ & 1\\
\hline
$h_u $ & $\mathbb{I}$ & $\square$ & 1/2 \\
\hline
$h_d $ & $\mathbb{I}$ & $\overline\square$ & -1/2 \\
\hline
$\lambda_3$ & Ad & $\mathbb{I}$ & 0 \\
\hline
$\lambda_2 $ & $\mathbb{I}$ & Ad & 0 \\
\hline
$\lambda_1 $ & $\mathbb{I}$ & $\mathbb{I}$ & 0\\
\hline
$H $ & $\mathbb{I}$ & $\square$ & 1/2\\
\hline
$Q$ & $\square$ & $\square$ & 1/6 \\
\hline
$U^c$ & $\overline\square$ & $\mathbb{I}$ & -2/3 \\
\end{tabular} 
\end{center}\par

As argued in ref.~\cite{Tamarit:2012}, the Lagrangian can be taken without loss of generality (assuming lepton number conservation) as
\begin{align}
\nonumber {\cal L}=&{\cal L}_{SM}\!-\Big\{\!\mu h_u h_d \!+\!{z_q}_j U^c q_j\epsilon h_u\!+\!{z_u}_jQ\epsilon h_u u^c_j+{z_d}_jQ h_d d^c_j+\frac{1}{2}\sum_{k=1}^3\sum_{A=1}^{l(k)}M_k \lambda^A_k\lambda^A_k\\
\nonumber&\left.+\sum_{k=1}^3\sum_{A=1}^{l(k)}(g_{H_k}H^\dagger T_k^A\lambda^A h_u+g_{H^*_k}H T_k^A\lambda^A h_d+g_{Q_{j,k}} Q^\dagger T_k^A\lambda^A q_j+g_{U_{j,k}} {U^c}^\dagger T_k^A\lambda^A u^c_j)+{\rm{c.c.}}\Big\}\right.\\
\nonumber&-\frac{1}{2}\sum_{k=1}^3\sum_{A=1}^{l(k)}\gamma_{k,S,S'}D^{k,A}_S D^{k,A}_{S'} -m^2_Q Q^\dagger Q-m^2_U {U^c}^\dagger U^c-(a_uQ\epsilon H U^c+{\rm c.c.}),\\
& D^{k,A}_S\equiv S^\dagger T_k^A S. \label{eq:LagrangianMES}
\end{align}
In the expressions above,  $S$ denotes the scalar fields in the theory, and $\gamma_{k,S,S'}=\gamma_{k,S',S}$; $j$ is summed over and runs over the three generations. Also, $q_j\epsilon h_u=q^a_j\epsilon^{ab} h^b_u$, (and similarly for the rest of the terms involving $\epsilon$), with  $a,b$ being SU(2) indices of the fundamental representation, and $\epsilon^{12}=1$. Given the transformation properties of the fields under the gauge groups, $g_{H_3}=g_{H_3^*}=g_{U_{j,2}}=\gamma_{2,S,U}=\gamma_{3,H,S}=0$. Moreover, in ref.~\cite{Tamarit:2012} it was shown that some of the quartic couplings are redundant; some of them have already been eliminated in eq.~\eqref{eq:LagrangianMES}, but one can impose as well
\begin{align*}
 \gamma_{2,H,H}=\gamma_{3,Q,Q}=\gamma_{3,U,U}=0.
\end{align*}
The fact that the resulting $\gamma$'s absorb other quartic couplings has to be taken into account when imposing SUSY boundary conditions at the scale of the heavy scalars; this is done in appendix \ref{app:highscale}.

\subsection{Effective SUSY with a full generation of light scalars\label{subsec:Lagrangians:NonMinEffSUSY}}

In this case the fields of the low energy theory are those in the previous sections plus a right-handed sbottom, a left-handed third-generation slepton doublet and a right handed stau, as summarized in the next table.

\begin{center}
\begin{tabular}{c|c|c|c}
 \   & SU(3) & SU(2) & U(1) \\
\hline  
$q_i,i=1\dots 3$ & $\square$ & $\square$ & 1/6 \\
\hline
$u^c_{i},i=1\dots 3$ & $\overline\square$ & $\mathbb{I}$ & -2/3 \\
\hline
$d^c_{i},i=1\dots 3$ & $\overline\square$ & $\mathbb{I}$ & 1/3 \\
\hline
$l_{i},i=1\dots 3$ & $\mathbb{I}$ & $\square$ & -1/2 \\
\hline
$e^c_{i},i=1\dots 3$ & $\mathbb{I}$ & $\mathbb{I}$ & 1\\

\end{tabular}

\begin{tabular}{c|c|c|c}
 \   & SU(3) & SU(2) & U(1) \\
\hline  
$h_u $ & $\mathbb{I}$ & $\square$ & 1/2 \\
\hline
$h_d $ & $\mathbb{I}$ & $\overline\square$ & -1/2 \\
\hline
$\lambda_3$ & Ad & $\mathbb{I}$ & 0 \\
\hline
$\lambda_2 $ & $\mathbb{I}$ & Ad & 0 \\
\hline
$\lambda_1 $ & $\mathbb{I}$ & $\mathbb{I}$ & 0\\
\hline
$H $ & $\mathbb{I}$ & $\square$ & 1/2\\
\hline
$Q$ & $\square$ & $\square$ & 1/6 \\
\hline
$U^c$ & $\overline\square$ & $\mathbb{I}$ & -2/3 \\
$D^c$ & $\overline\square$ & $\mathbb{I}$ & 1/3 \\
\hline
$L$ & $\mathbb{I}$ & $\square$ & -1/2 \\
\hline
$E^c$ & $\mathbb{I}$ & $\mathbb{I}$ & 1
\end{tabular} 
\end{center}\par
Without loss of generality, assuming lepton number conservation and eliminating redundant quartic couplings, the Lagrangian can be written as \cite{Tamarit:2012}
\begin{align}
\nonumber {\cal L}=&{\cal L}_{SM}\!-\left\{\!\mu h_u h_d +{z_q}_j U^c q_j\epsilon h_u+{z_u}_jQ \epsilon h_u u^c_j+{z_d}_jQ h_d d^c_j+{z_{q^*}}_jD^c q_jh_d+{z_{l}}_jE^c l_jh_d\right.\\
\nonumber&\left.+{z_{e}}_jL e^c_jh_d+\frac{1}{2}\sum_{k=1}^3\sum_{A=1}^{l(k)}M_k \lambda^A_k\lambda^A_k+\sum_{k=1}^3\sum_{A=1}^{l(k)}(g_{H_k}H^\dagger T_k^A\lambda^A h_u+g_{H^*_k}H T_k^A\lambda^A h_d\right.\\
\nonumber&+g_{Q_{j,k}} Q^\dagger T_k^A\lambda^A q_j+g_{U_{j,k}} {U^c}^\dagger T_k^A\lambda^A u^c_j+g_{D_{j,k}} {D^c}^\dagger T_k^A\lambda^A d^c_j+g_{L_{j,k}} {L}^\dagger T_k^A\lambda^A l_j\\
\nonumber&\left.+g_{E_{j,k}} {E^c}^\dagger T_k^A\lambda^A e^c_j)+\rm{c.c.}\right\}-\frac{1}{2}\sum_{k=1}^3\sum_{A=1}^{l(k)}\gamma_{k,S,S'}D^{k,A}_S D^{k,A}_{S'}-m^2_Q Q^\dagger Q-m^2_U {U^c}^\dagger U^c\\
&\nonumber-m^2_D {D^c}^\dagger D^c-m^2_L L^\dagger L
-m^2_E {E^c}^\dagger E^c
-(a_u Q\epsilon H U^c+c_d Q H^\dagger D^c+c_l Q H^\dagger E^c\\
&+\lambda'_{E}(Q L^\dagger)({E^c}^\dagger D^c)+\lambda''_{E}(Q\epsilon L)(E^cU^c)+{\rm c.c.}), \label{eq:LagrangianNMES}
\end{align}
In this case, given the properties of the fields under gauge transformations, $g_{H_3}=g_{H_3^*}=g_{U_{j,2}}=g_{D_{j,2}}=g_{L_{j,3}}=g_{E_{j,2}}=g_{E_{j,3}}=\gamma_{2,U/D/E,S}=\gamma_{3,H/L/E,S}=0$; the notation and summing conventions are the same as in eq.~\eqref{eq:LagrangianMES}. Again, some of the $\gamma_{k,S,S'}$ are redundant and can be taken to zero; a possible choice is \cite{Tamarit:2012}
\begin{align*}
  \gamma_{2,H,H}=\gamma_{2,L,L}=\gamma_{3,Q,Q}=\gamma_{3,U,U}=\gamma_{3,D,D}=0.
\end{align*}
The resulting $\gamma$'s absorb some of the other quartic couplings, which has to be taken into account when imposing supersymmetric boundary conditions at the scale of the heavy scalars, as is explained in \S \ref{app:highscale}.

\section{Gauge coupling unification\label{sec:Unification}}

In order to study gauge coupling unification, the two-loop RG equations for dimensionless parameters are run in three stages across two thresholds, $\Lambda_{SM}$ and $\Lambda$, marking the range of validity of the effective theories in the previous section. The running ignores off-diagonal flavor contributions and phases. First, PDG Data \cite{Nakamura:2012} are used to compute the Standard Model gauge couplings in the $\overline{\rm MS}$ scheme at the scale of the Z boson's mass; next, they are evolved with the 2-loop SM RG equations \cite{Luo:2002ey} up to the scale of the top mass, using  tree-level estimates for the initial values of the top and bottom Yukawas and a guessed value for the Higgs quartic coupling. At $q=m_t$, the values of  $y_t$ and $y_b$ are recalculated from the pole masses of the quarks by applying QCD threshold corrections of order $\alpha_s^2$  --see eqs.~\ref{eq:QCDthr0} and \ref{eq:QCDthr}. The two-loop SM RG is continued up to the  Effective SUSY threshold $\Lambda_{SM}$, around a 
TeV. There the SM couplings are matched with the Effective SUSY ones, including one-loop threshold corrections for the Yukawa and gauge couplings, as explained in \S \ref{app:lowscale}, see eqs.~\eqref{eq:MESgthr}-\eqref{eq:nMESythr}. Guess values are given to the rest of the parameters, which are evolved at two-loops till the MSSM threshold $\Lambda$, of the order of 10-20 TeV. At this scale, supersymmetric boundary conditions taking into account one-loop $\overline{\rm DR}$ to $\overline{\rm MS}$ conversion factors are enforced for the Yukawa and quartic couplings, as detailed in \S \ref{subapp:matchhigh}.

Following this, the RG equations are run recursively up and down between $m_t$ and the MSSM threshold $\Lambda$, applying the appropriate boundary conditions at each endpoint and doing the correct matching for all couplings across the Effective SUSY threshold $\Lambda_{SM}$, until convergence is achieved between the susy boundary conditions at the MSSM scale and the values obtained from the upwards RG evolution. In the end, couplings are matched across the MSSM threshold as described in \S \ref{app:highscale} and the MSSM ${\overline{\rm DR}}$ equations are run towards high scales.

The free parameters are the two threshold scales, the Higgs mixing angle $\alpha$ entering the susy boundary conditions\footnote{$\alpha$ is related to the MSSM parameter $\beta$ in the Higgs decoupling limit by $\tan\beta\sim\cot\alpha$, see  \S \ref{subapp:higgsdecoupling}.} from the relation
\begin{align*}
H=\cos\alpha H_u+\sin\alpha H_d^\dagger,
\end{align*}
and the supersymmetric fermion and scalar masses and the mixing angles entering the threshold corrections for the gauge and Yukawa couplings.

\begin{figure}[h]\centering
  \includegraphics[scale=1.5]{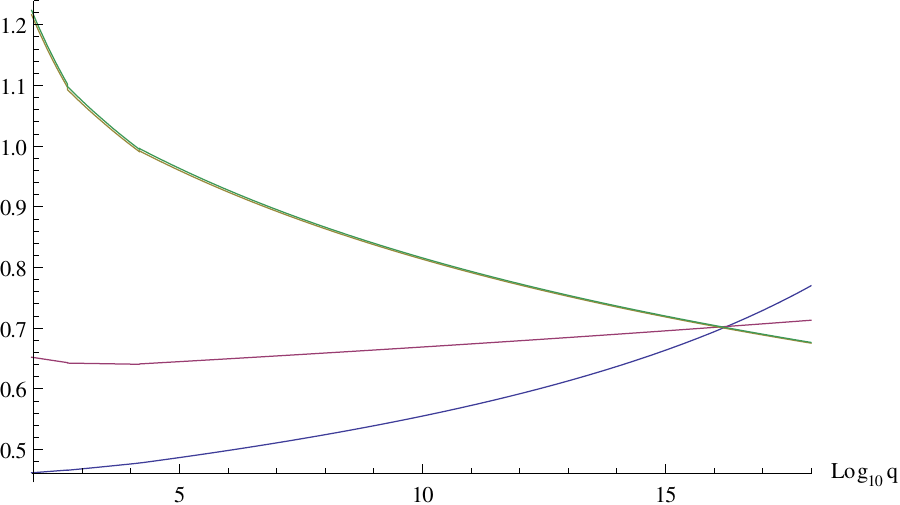}
\caption{\label{fig:1}2 loop running of the gauge couplings in a nonminimal Effective SUSY scenario, with thresholds at 0.5 and 15 TeV, $\tan\beta\sim\cot\alpha=10$. Fermions are assumed to lie around 0.5 TeV, and scalars at 1 TeV.}
\end{figure}

Fig.~\ref{fig:1} shows an example of the running of the gauge couplings with scale in a nonminimal Effective SUSY scenario for $\cot\alpha=10$, with the thresholds at 0.5 and 15 TeV, the supersymmetric fermions at around 0.5 TeV and the scalars at around 1 TeV. The couplings unify within 0.5\% precision.

\begin{figure}[h!]\centering
\begin{minipage}{0.5\textwidth}\centering
   \includegraphics[scale=0.8]{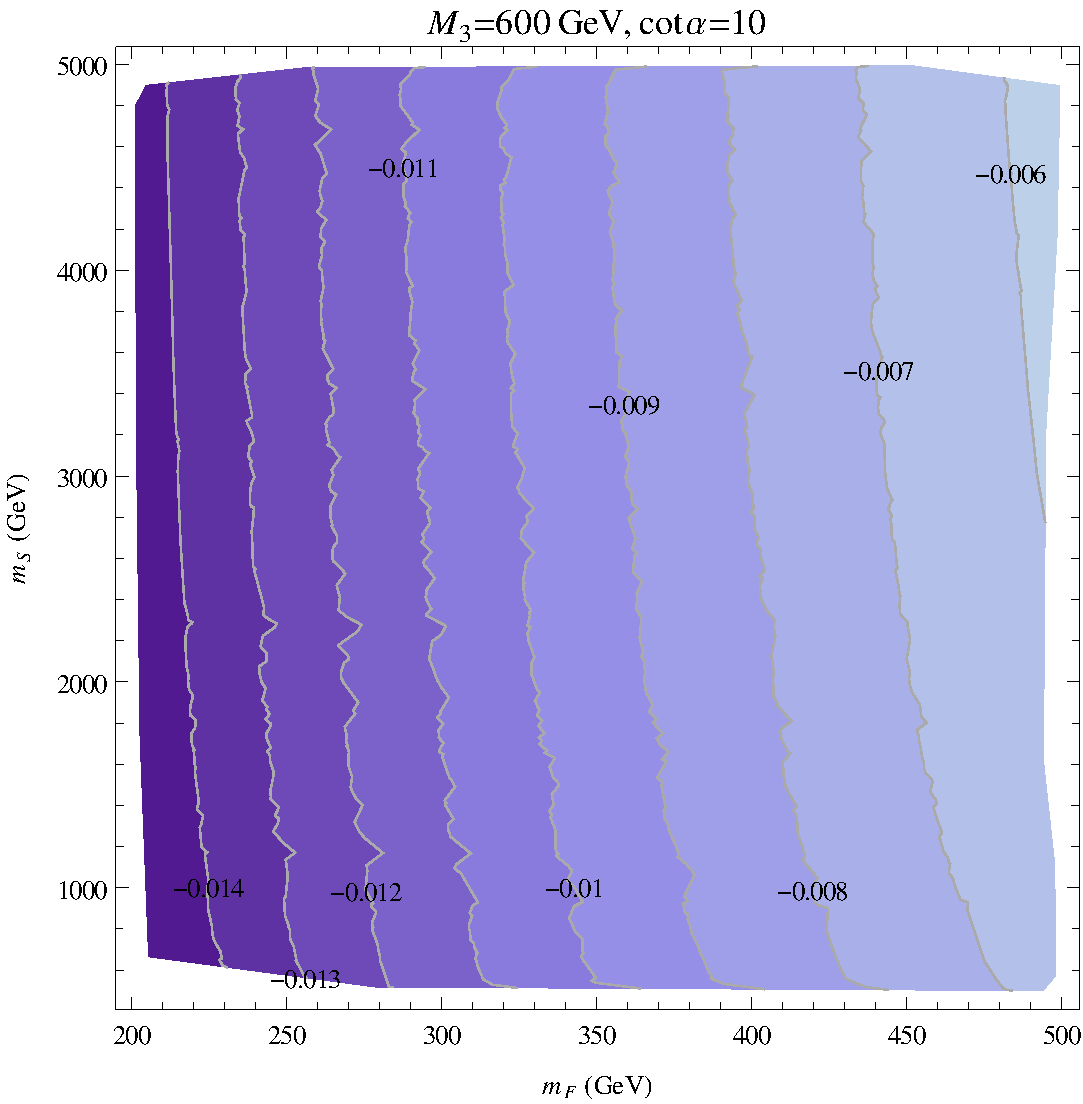}
\end{minipage}\begin{minipage}{0.5\textwidth}\centering
 \includegraphics[scale=0.8]{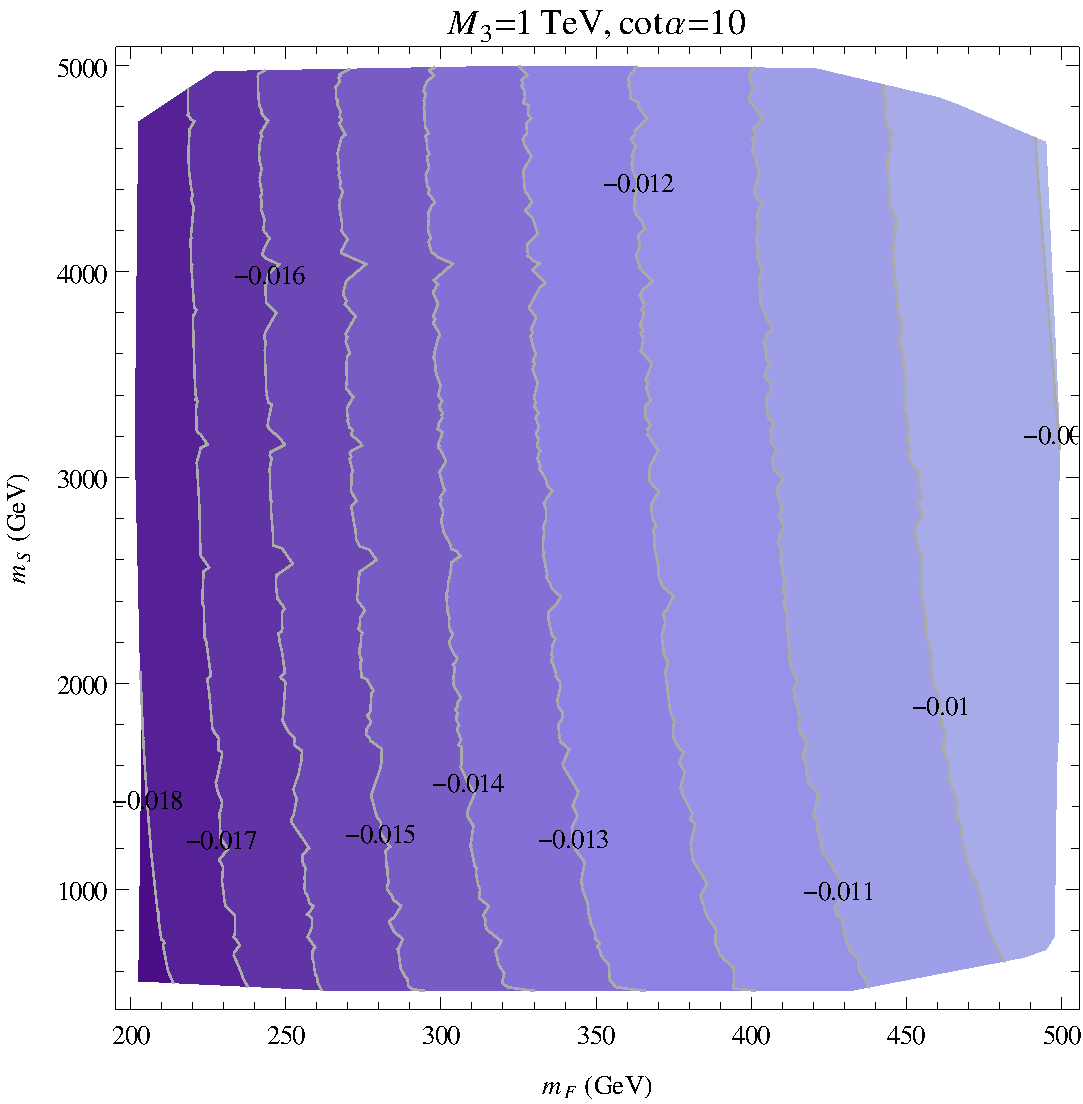}
\end{minipage}
\caption{\label{fig:2} Values of the unification scale threshold parameter $\epsilon_g$ in minimal Effective SUSY scenarios, for $\cot\alpha=10$ and two choices of $M_3$, in terms of a common fermion mass $m_F$ and a common scalar mass $m_S$ used to calculate threshold corrections.}
\end{figure}

\begin{figure}[h!]\centering
\begin{minipage}{0.5\textwidth}\centering
   \includegraphics[scale=0.8]{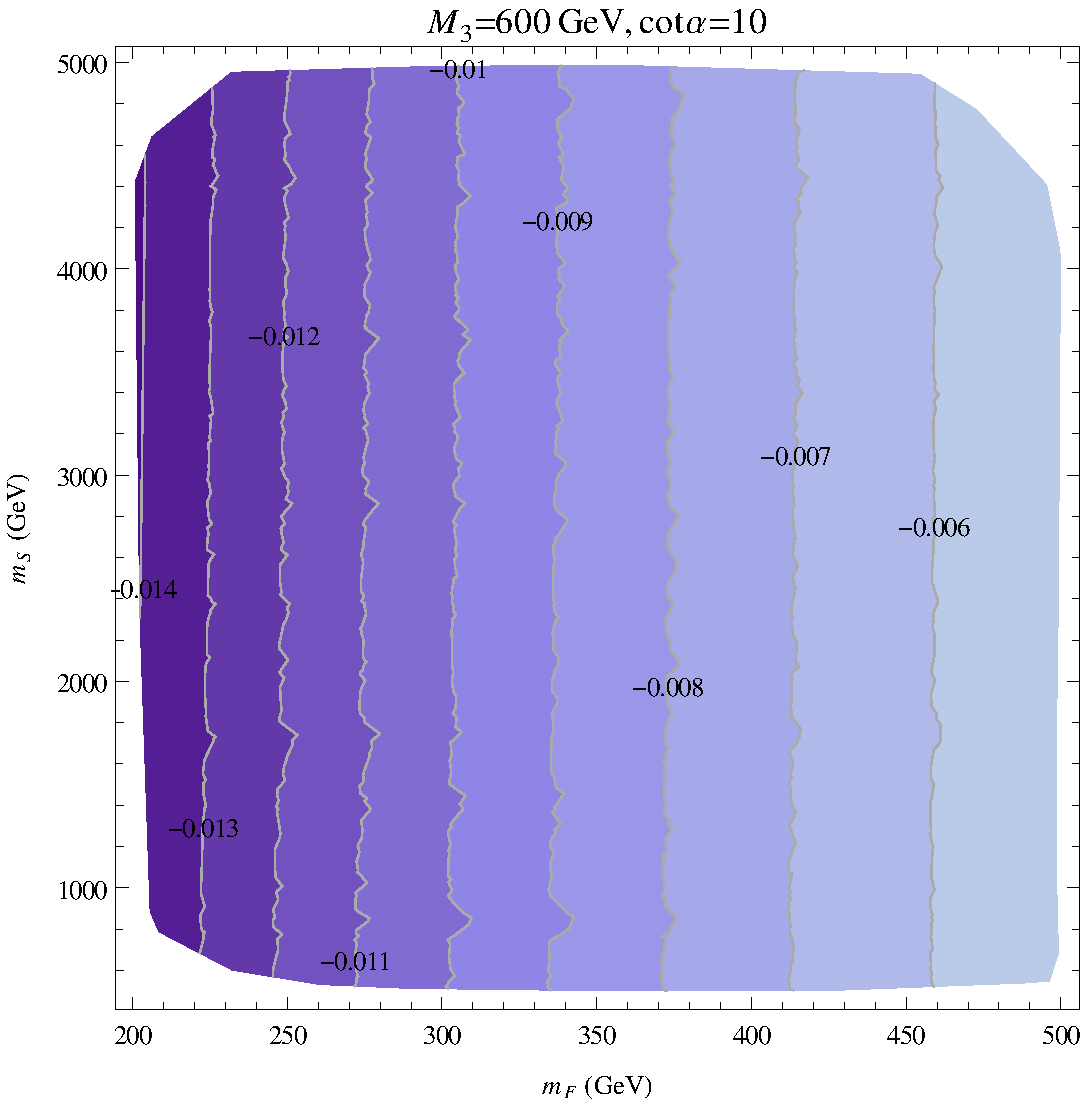}
\end{minipage}\begin{minipage}{0.5\textwidth}\centering
 \includegraphics[scale=0.8]{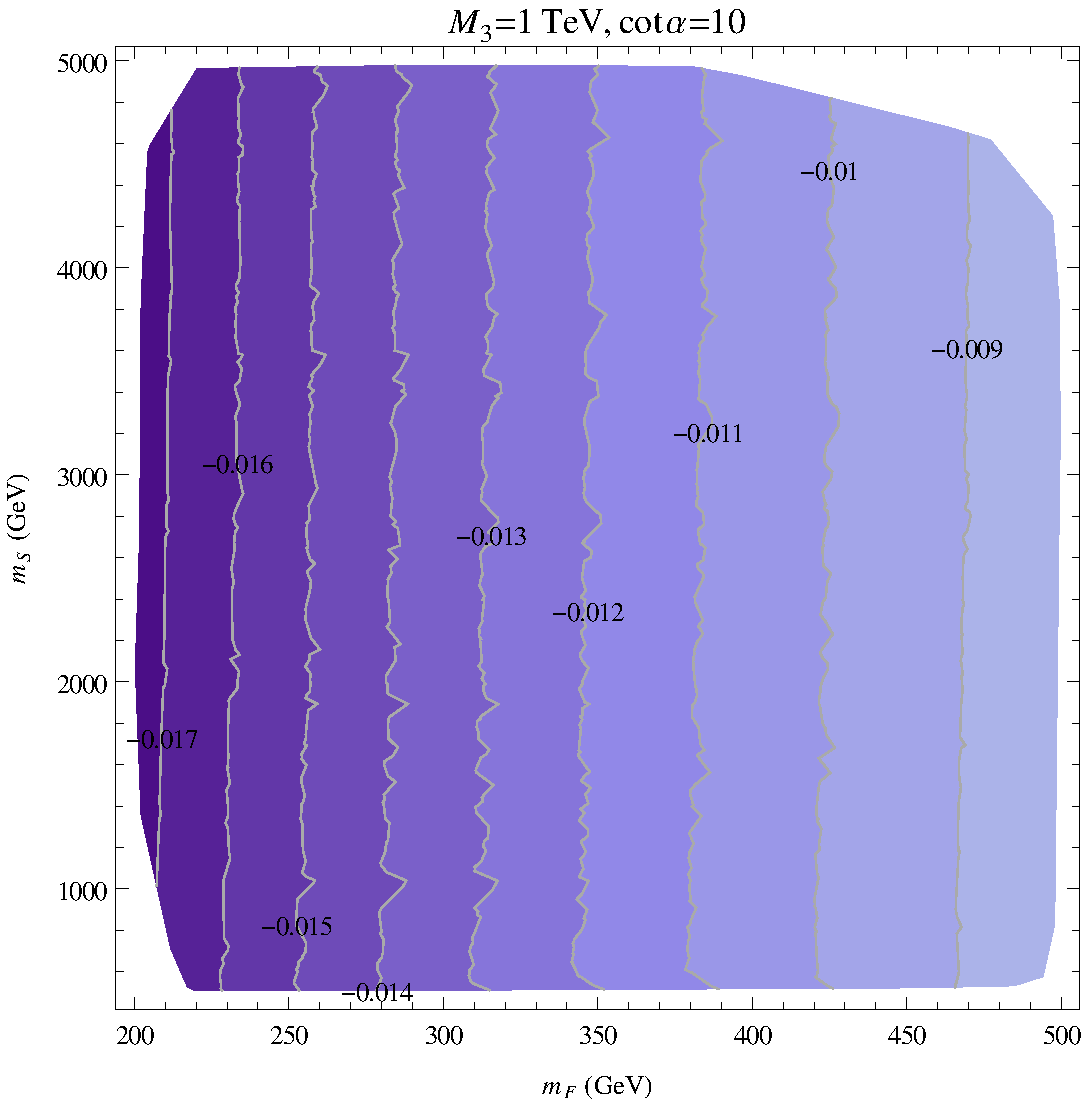}
\end{minipage}
\caption{\label{fig:3} 
Values of the unification scale threshold parameter $\epsilon_g$ in nonminimal Effective SUSY scenarios, for $\cot\alpha=10$ and two choices of $M_3$, in terms of a common fermion mass $m_F$ and a common scalar mass $m_S$ used to calculate threshold corrections. $c_d,c_l$ were set to zero at the high scale.}
\end{figure}

In order to quantify the precision of unification and compare it with that in standard MSSM  scenarios, one can define a threshold parameter $\epsilon_g$  as \cite{Pierce:1996zz}
\begin{align}\label{eq:epsilong}
g_3(q_{\rm GUT})=g_2(q_{\rm GUT})(1+\epsilon_g),
\end{align}
where $q_{\rm GUT}$ is the scale at which $g_1$ and $g_2$ meet. Fig.~\ref{fig:2} shows the values of $\epsilon_g$ in minimal Effective SUSY scenarios for different values of $\cot\alpha$ and the gluino mass $M_3$, as a function of fermion and scalar masses $m_F$, $m_S$ which are used to set the values of the rest of the dimensionful parameters. Fig.~\ref{fig:3} shows the corresponding results in the nonminimal scenario, when $c_d, c_l$ are set to zero at the high scale.

\begin{figure}[h!]\centering
\begin{minipage}{.5\textwidth}\centering
   \includegraphics[scale=.9]{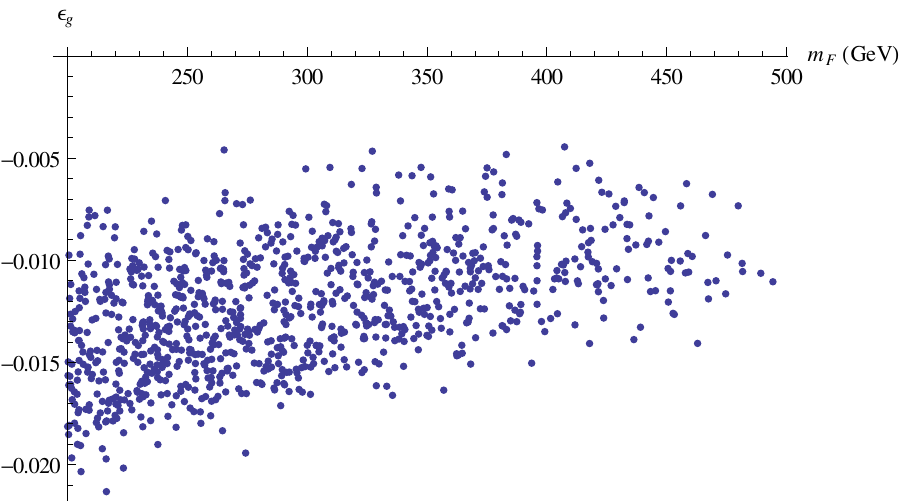}
\end{minipage}\begin{minipage}{0.5\textwidth}\centering
 \includegraphics[scale=.9]{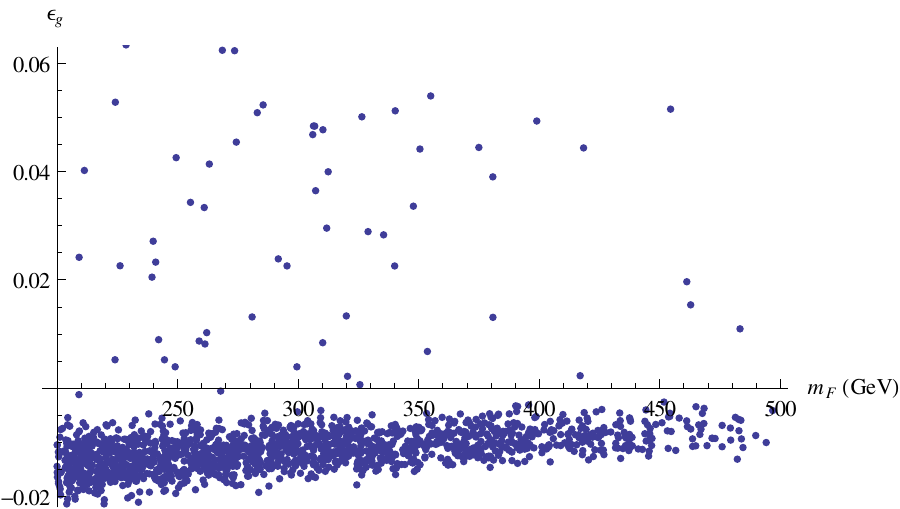}
\end{minipage}
\begin{minipage}{0.5\textwidth}\centering
   \includegraphics[scale=.9]{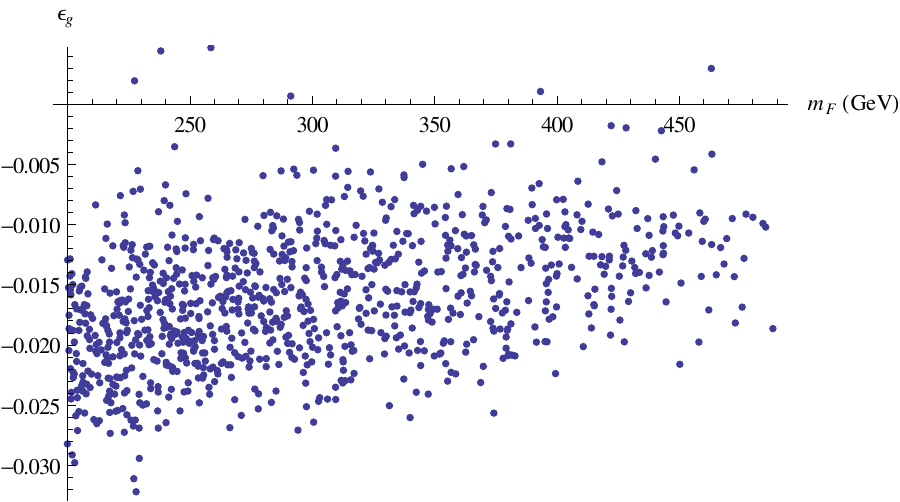}
\end{minipage}\begin{minipage}{0.5\textwidth}\centering
   \includegraphics[scale=.9]{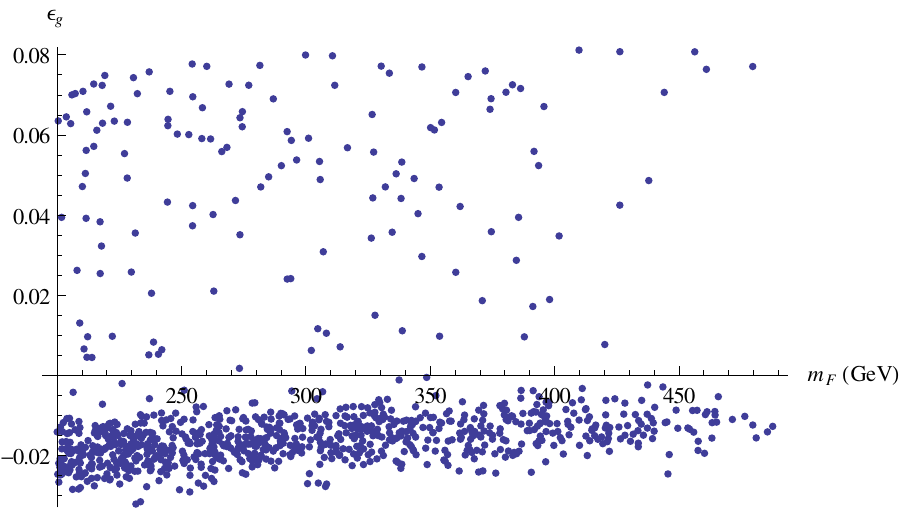}
\end{minipage}
\caption{\label{fig:unifscansall} Values of the unification scale threshold parameter $\epsilon_g$ in minimal Effective SUSY scenarios (upper left), nonminimal ones (upper right), and ``standard'' MSSM scenarios with $a_d=0$ at the high scale, focusing on the region $\epsilon<0.005$ (lower left), and allowing for large sbottom mixing (lower right), as a function of the minimum mass scale of the fermionic superpartners.}
\end{figure}

 It is clear from figs.~\ref{fig:2} and \ref{fig:3} that in the limit of small $c_d,c_l$ the results are similar for both Effective SUSY scenarios, and the value of $\epsilon_g$ is mostly influenced by the mass of higgsinos and light gauginos. The values of $\epsilon_g$ are slightly lower than in standard MSSM scenarios, in which they are usually between $-1\%$ and $-3\%$ \cite{Pierce:1996zz}. Concerning nonzero values of $c_d, c_l$, they can give rise to positive values of $\epsilon_g$ due to large sbottom mixing angles enhancing the bottom Yukawa through the threshold corrections of eq.~\eqref{eq:nMESythr}. In order to facilitate direct comparison, Fig.~\ref{fig:unifscansall} shows the values of $\epsilon_g$ as a function of the lightest fermion mass parameters in the two Effective SUSY scenarios and in the standard MSSM case, obtained from  scans in which $\cot\alpha$ takes random values between 5 and 60, while the mass parameters entering threshold corrections are varied randomly in the following windows:
 $|\mu|,|M_2|$  between 200 and 500 GeV, $|M_3|$ between 500 and 1500 GeV, and dimensionful scalar parameters between 500 and 4500 GeV (allowing for negative values of the trilinear couplings). Note that, as anticipated, there are positive values of $\epsilon_g$ in nonminimal Effective SUSY scenarios, as well as in the standard MSSM. It should be noted that positive values of $\epsilon$ become more seldom  when rescaling the trilinear couplings in proportion to the corresponding Yukawas, as is usually done in the literature, so that much lower values of $c_d$ or the MSSM $a_d$ coupling are probed, as in ref.~\cite{Pierce:1996zz}. For this reason Fig.~\ref{fig:unifscansall} includes MSSM results with $a_d,c_d$ set to zero at the high scale and focusing on the region $\epsilon<0.005$ (See eq.~\eqref{eq:MSSMaterms} for the notation concerning the MSSM trilinear couplings).

Regarding uncertainties in the results, all the figures for $\epsilon_g$ were obtained using the central value for $\alpha_s=0.1184(7)$, and changing $\alpha_s$ within its error gives rise to changes in $\epsilon_g$ of around $\pm 0.0009$. 

\section{$b$-$\tau$ unification\label{sec:btauunif}}

\begin{figure}[h!]\centering
\begin{minipage}{.5\textwidth}\centering
   \includegraphics[scale=.9]{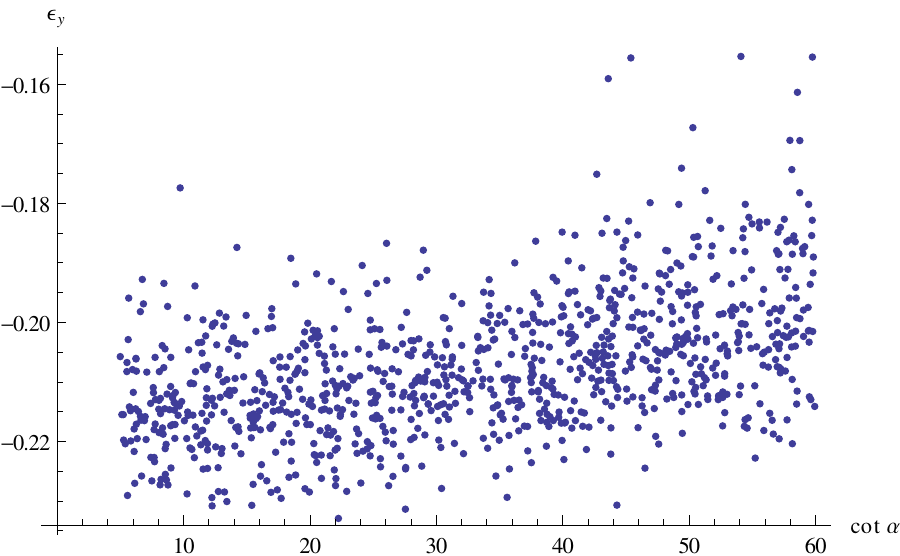}
\end{minipage}\begin{minipage}{0.5\textwidth}\centering
 \includegraphics[scale=.9]{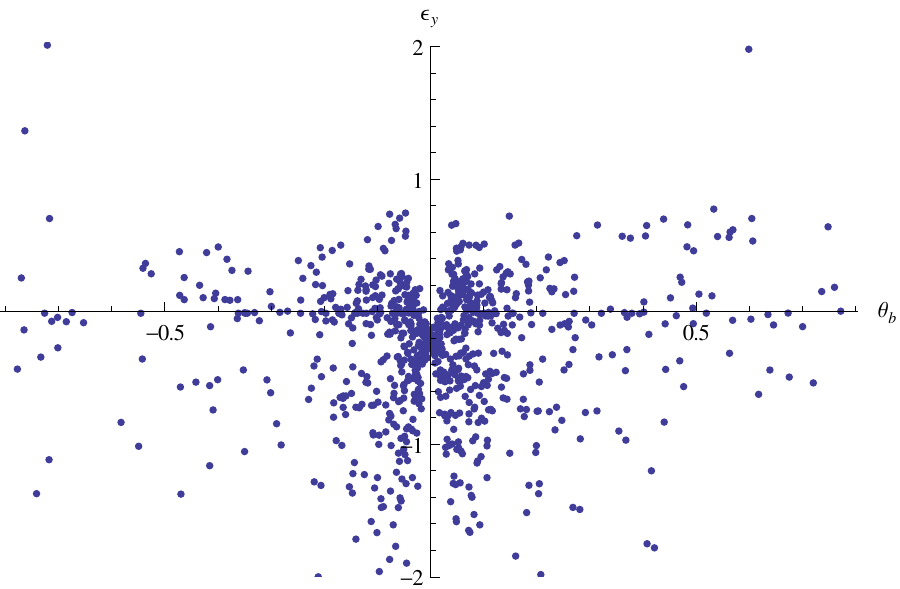}
\end{minipage}
\begin{minipage}{0.5\textwidth}\centering
   \includegraphics[scale=.9]{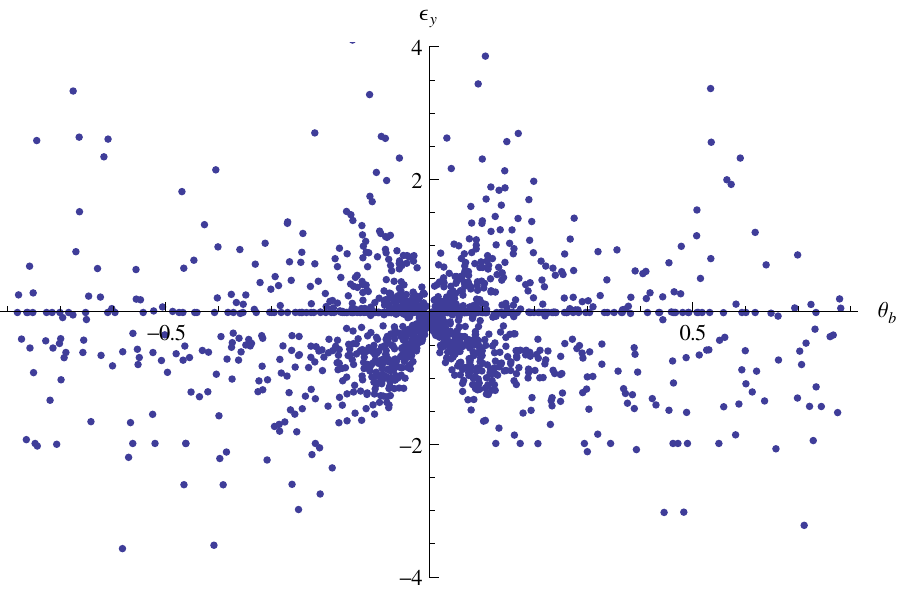}
\end{minipage}
\caption{\label{fig:btau} Upper left: Values of the unification scale threshold parameter $\epsilon_y$ in minimal Effective SUSY scenarios versus $\cot\alpha$. Upper right: $\epsilon_y$ versus the sbottom mixing angle $\th_b$ in nonminimal scenarios. Bottom:  $\epsilon_y$  versus $\th_b$  in``standard'' MSSM scenarios.}
\end{figure}

After having solved the RG equations for the dimensionless parameters with appropriate boundary conditions at high and low scales as explained in the previous section, one may also look into the unification of the bottom and tau Yukawa couplings, which is necessary for SU(5)-type models in which a single Higgs field gives masses to the matter fields. Defining a threshold parameter for the bottom and tau Yukawas as
\begin{equation*}
 \epsilon_y=\frac{y_b(q_{GUT})-y_\tau(q_{GUT})}{y_\tau(q_{GUT})},
\end{equation*}
and performing scans for $\cot\alpha$ between 5 and 60, $|\mu|,|M_2|$  between 200 and 500 GeV, $|M_3|$ between 500 and 1500 GeV, and dimensionful scalar parameters between 500 and 4500 GeV (again allowing for negative values of the trilinear couplings), one obtains the results of Fig.~\ref{fig:btau}.

In minimal Effective SUSY scenarios, $b$-$\tau$ unification seems  challenging; the bottom and Yukawa couplings tend to cross at too small scales. In nonminimal scenarios, the range for possible values of $ \epsilon_y$ is much larger, including zero, and similar to that in standard MSSM scenarios. The difference is due to the the nonzero sbottom mixing angle in the latter two cases, which greatly affects the threshold corrections for the bottom Yukawa (see eq.~\eqref{eq:nMESythr}). It should be pointed out however that the precision of the values of $\epsilon_y$ is rather small, mainly due to the errors in the bottom mass. All the figures for $\epsilon_y$ were obtained using the central values $\alpha_s=0.1184(7)$ and the pole bottom quark mass $m_b=4.78^{+0.20}_{-.07}$. Changing them within their uncertainty gives rise to changes in $\epsilon_y$ of up to  $\pm 0.05$ in the minimal Effective SUSY scenarios; the variations in the nonminimal Effective SUSY and standard MSSM scenarios can be much larger, up to 
$\pm 0.27$ and $\pm 0.7$, respectively.


\section{Higgs masses at two loops\label{sec:Higgs}}

The two-loop RG equations for the dimensionful parameters in Effective SUSY scenarios allow to calculate the particle spectrum. The Higgs boson is expected among the lightest particles, so that it becomes relevant to analyze the allowed values for its mass, more so given the current experimental hints for a 125-126 GeV particle. Another question worth studying is the degree of fine-tuning of possible Effective SUSY scenarios with a 125 GeV Higgs; this will be analyzed in the next section.

The results for Higgs masses in Effective SUSY scenarios presented here have been computed by using the full one-loop effective potential plus two-loop scalar corrections. Some details about the effective potential are given in \S \ref{app:VCW}. The computation requires to solve RG equations including massive parameters; boundary conditions for these have to be set at a given scale, which is chosen as the cutoff  scale $\Lambda$ of the heavy sparticles. There is yet a subtlety, which is that requiring a successful electroweak symmetry giving rise to the correct Higgs vacuum expectation value imposes a constraint in the high scale parameters via the minimization condition for the effective potential; it was chosen to determine $m^2_H$ in terms of other massive parameters.

The computation of the full RG flow in a manner consistent with electroweak symmetry breaking proceeds then as follows. First, the equations for the dimensionless couplings are integrated as explained in \S~\ref{sec:Unification}. The massive parameters are then run down from their boundary values at the scale $\Lambda$ till the lower Effective SUSY threshold $\Lambda_{SM}$, using a guessed boundary value for $m^2_H$.  At the scale  $\Lambda_{SM}$, the minimization condition for the effective potential is used to solve for $m^2_H$ in terms of the other massive parameters, fixing the Higgs vacuum expectation value (VEV) at the value determined by the experiments. With this value of $m^2_H$, the RG equations are run upwards again, high scale boundary conditions imposed without modifying $m^2_H$, and the process is repeated until convergence between successive determinations of $m^2_H$ at the lower threshold is achieved within a given tolerance. Again, flavour mixing effects and complex phases are ignored in the analysis.

\begin{figure}[h!]\centering
\begin{minipage}{.5\textwidth}\centering
   \includegraphics[scale=.9]{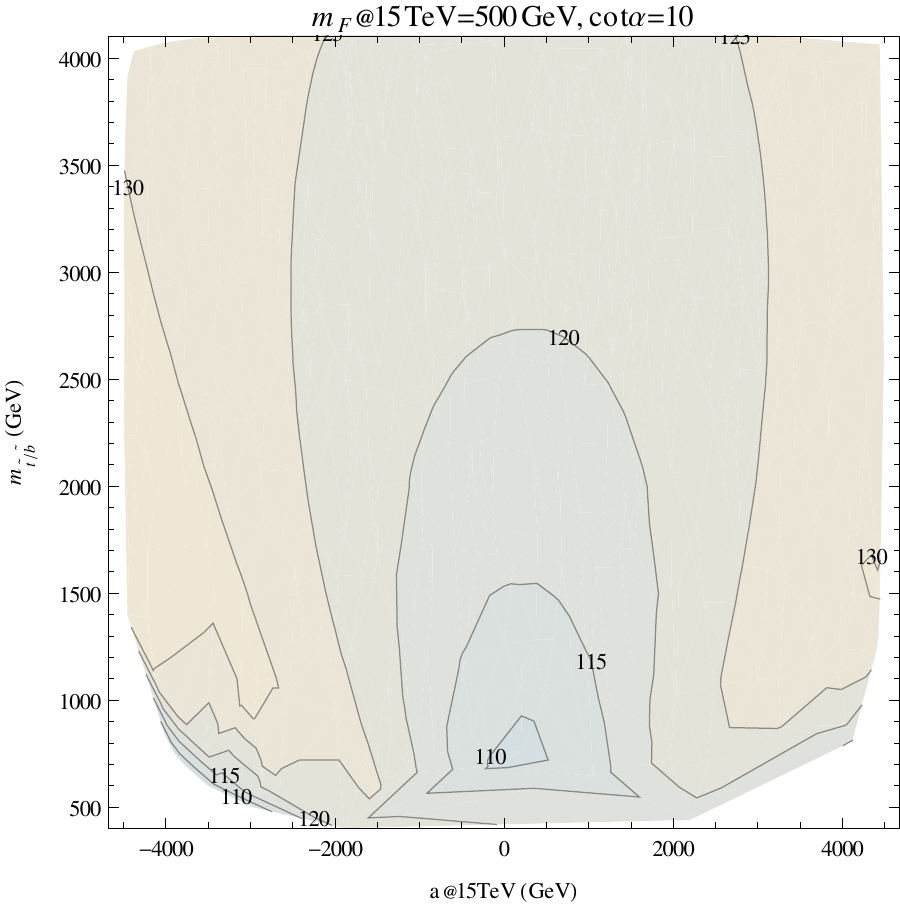}
\end{minipage}\begin{minipage}{.5\textwidth}\centering
   \includegraphics[scale=.9]{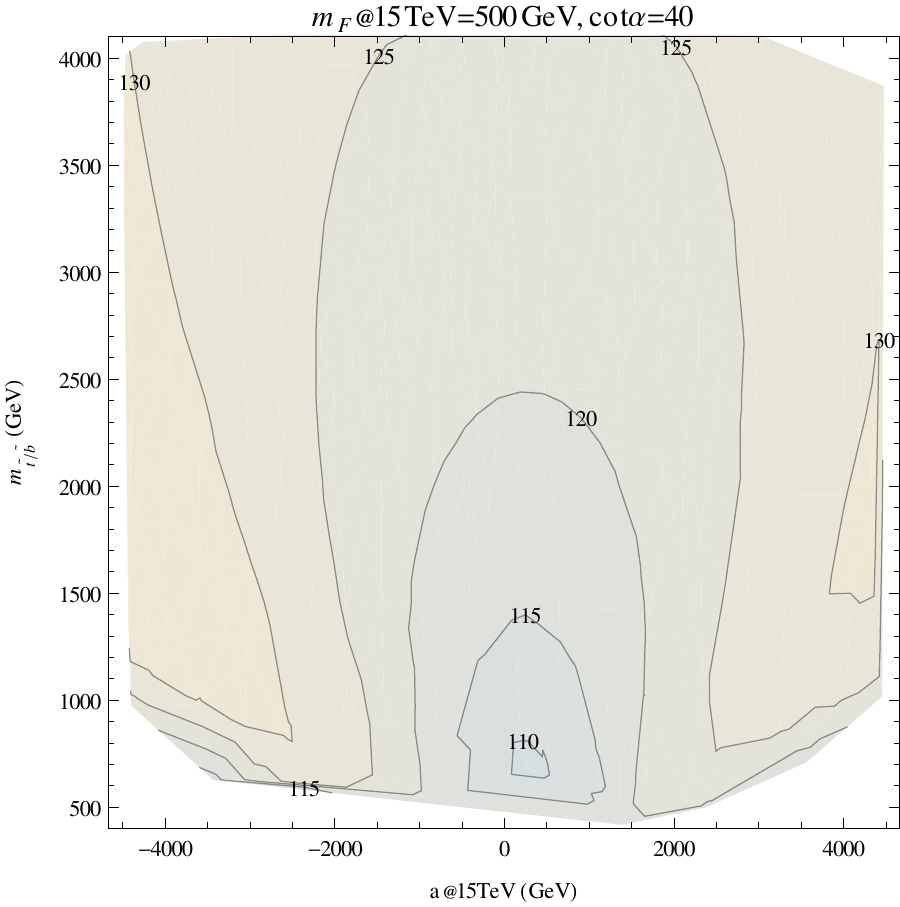}
\end{minipage}
\caption{\label{fig:scan2dMES} Higgs masses computed from the effective potential in minimal Effective SUSY scenarios for 2 different values of $\cot\alpha$, in terms of the high scale boundary value for $a_u$ and the lightest colored scalar eigenvalue. Fermion mass parameters were fixed at 500 GeV in the boundary; common boundary values for the scalar soft parameters were assumed, $|a_u|^2=m^2_{Q}=m^2_U$.}
\end{figure}

\begin{figure}[h!]\centering
\begin{minipage}{.5\textwidth}\centering
   \includegraphics[scale=.9]{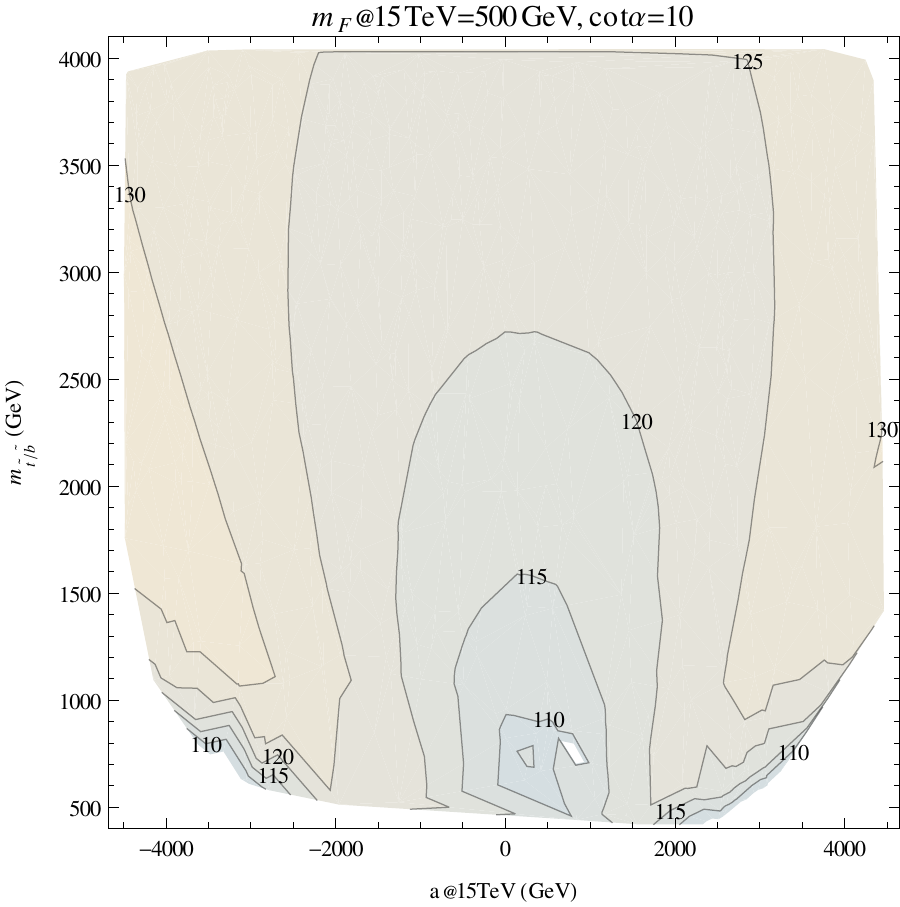}
\end{minipage}\begin{minipage}{.5\textwidth}\centering
   \includegraphics[scale=.9]{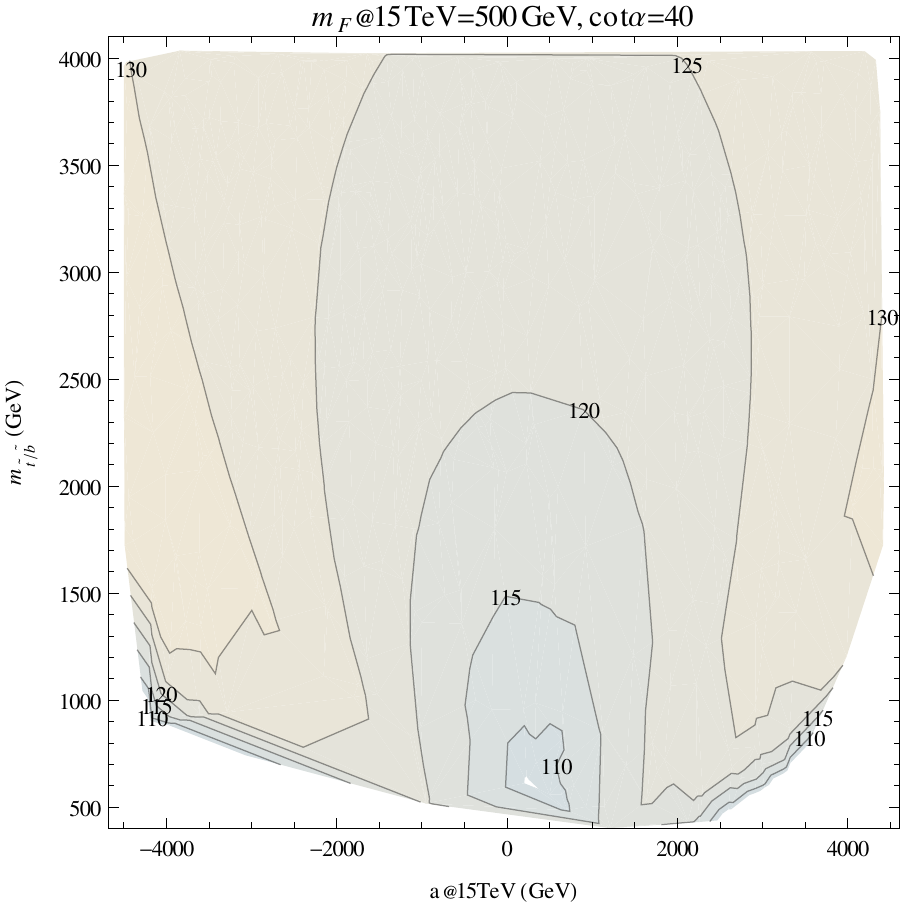}
\end{minipage}
\caption{\label{fig:scan2dES} Higgs masses computed from the effective potential in nonminimal Effective SUSY scenarios for 4 different values of $\cot\alpha$, in terms of the high scale boundary value of $a_u$, with $|a_u|^2=m^2_{Q}=m^2_U=m^2_D=m^2_L=m^2_E$, $c_d=c_l=0$ at the high scale. Fermion mass parameters were fixed at 500 GeV in the boundary.}
\end{figure}

Several scans over the parameter spaces were performed in order to identify the allowed range for the Higgs mass. First, two-dimensional scans were done by fixing the fermion masses and varying the scalar soft parameters, assuming degeneracies of the latter at the high scale safe for a possible factor of $\pm1$ for the trilinear couplings ($|a_u|^2=m^2_{Q}=m^2_U$); this was repeated for different values of $\cot\alpha\sim\tan\beta$. Fig.~\ref{fig:scan2dMES} shows the results of two of this scans for two different values of $\cot\alpha$ in minimal Effective SUSY scenarios, with fermion masses fixed at 500 GeV at the high scale. Higgs mass contours are given in terms of the boundary value for the $a_u$ term at the high scale threshold of 15 TeV and the mass of the lightest colored scalar eigenvalue $m_{\tilde t/\tilde b}$. The latter was computed including one-loop self-energy corrections  (adapted from ref.~\cite{Pierce:1996zz}, see eqs.~\ref{eq:mstop1} and \eqref{eq:mstop2}) at the lowest threshold, taken as 500 GeV. 
Fig.~\ref{fig:scan2dES} shows the results of analogous computations in 
nonminimal Effective SUSY scenarios, assuming again common values for the scalar soft parameters at 15 TeV (up to a sign in the trilinear coupling $a_u$), and taking $c_d=c_l=0$ at the high scale. 


\begin{figure}[h!]\centering
\begin{minipage}{.5\textwidth}\centering
   \includegraphics[scale=.9]{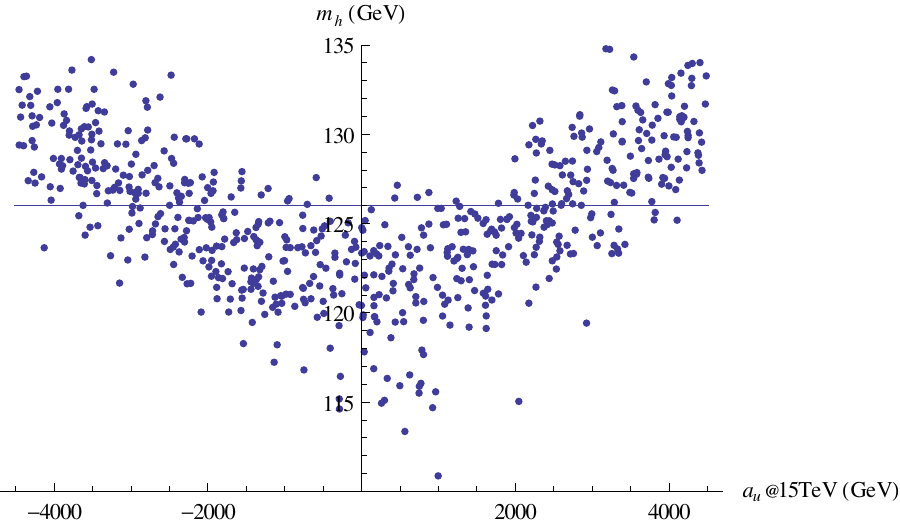}
\end{minipage}\begin{minipage}{.5\textwidth}\centering
   \includegraphics[scale=.9]{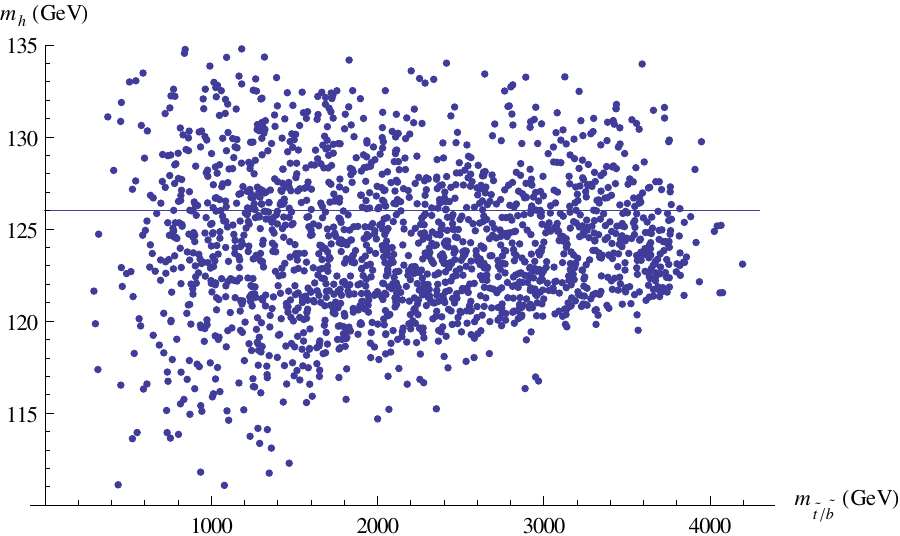}
\end{minipage}
\caption{\label{fig:scansMES} Higgs mass versus  the high scale value of the trilinear coupling $a_u$ (left), and the mass of the lightest colored scalar (right), in minimal Effective SUSY scenarios.}
\end{figure}

\begin{figure}[h!]\centering
\begin{minipage}{.5\textwidth}\centering
   \includegraphics[scale=.9]{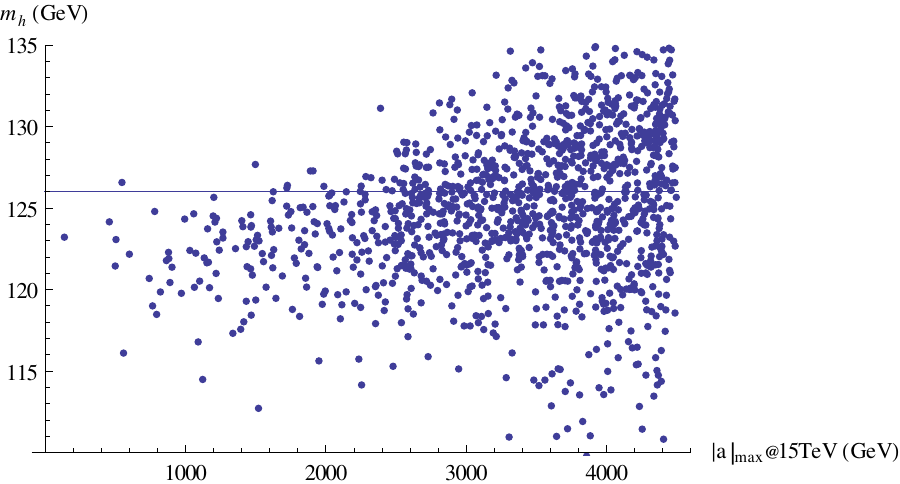}
\end{minipage}\begin{minipage}{.5\textwidth}\centering
   \includegraphics[scale=.9]{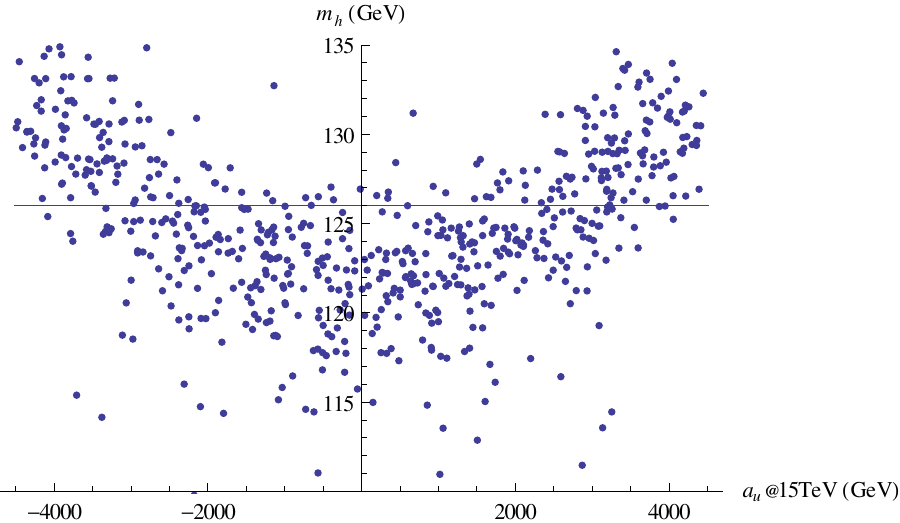}
\end{minipage}
\begin{minipage}{.5\textwidth}\centering
   \includegraphics[scale=.9]{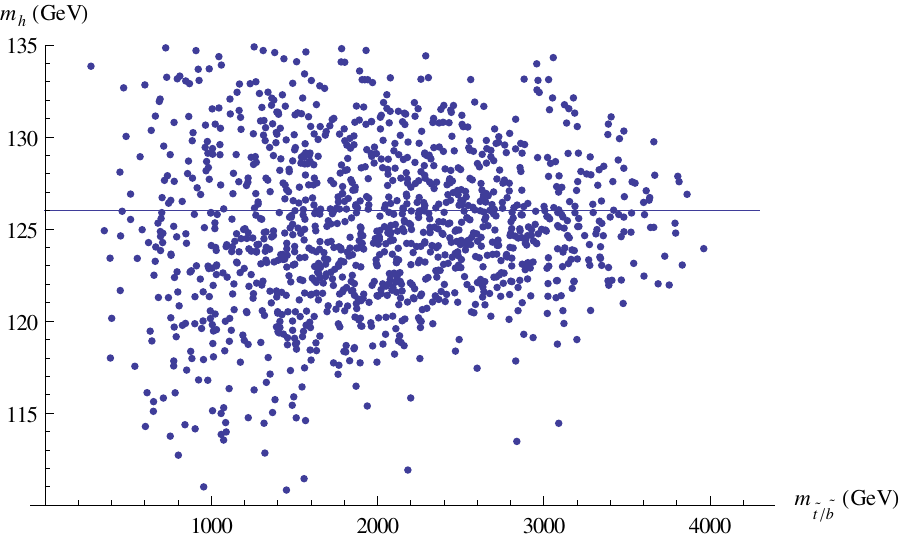}
\end{minipage}
\caption{\label{fig:scansES} Higgs mass in nonminimal Effective SUSY scenarios versus the maximum high scale value of the trilinear couplings $a_u,c_d,c_l$ (upper left), the high scale value of $a_u$ (upper right) and the mass of the lightest colored scalar.}
\end{figure}

Figs.~\ref{fig:scan2dMES} and \ref{fig:scan2dES} suggest that a 125-126 GeV Higgs with light sbottom/stops requires large trilinear couplings of around 2-3 TeV, and that the lightest colored superpartners in this case sit around 1 TeV. However, those figures were obtained assuming degeneracy of the scalar soft parameters at 15 TeV. In order to get more general results, scans were performed assigning random values to all dimensionful parameters within predefined intervals: $200$-$1000$ GeV for $|\mu|,|M_1|,|M_2|$, $200$-$1500$ GeV for $|M_3|$, $500^2$-$4500^2\,{\rm GeV}^2$ for scalar soft masses squared, and $500$-$4500$ GeV for the absolute value of the trilinear couplings. $\tan\beta$ was varied as well between 5 and 60. Fig.~\ref{fig:scansMES} shows the resulting Higgs masses in minimal Effective SUSY scenarios versus the trilinear coupling $a_u$ and the mass of the lightest colored particle. Fig.~\ref{fig:scansES} shows the corresponding results in nonminimal scenarios, plotted against the maximum size of the 
trilinear couplings $a_u,c_d,c_l$ at the high 
scale, against $a_u$ alone and versus the mass of the lightest colored particle. The calculations confirm the need of large trilinear couplings (around 2 TeV) in typical scenarios in order to get a Higgs mass of 125 GeV; it is interesting to note that the first two plots in Fig.~\ref{fig:scansES} suggest that one may have such a Higgs with small $a_u$ when the other trilinears get large. Allowing for non-degeneracies between the soft masses at the high scale makes it possible to have a 125 GeV Higgs with stops/bottoms around 300 GeV. In this region of parameter space, however, the uncertainty in the Higgs masses is higher, since the one-loop and two-loop results can differ in a couple of GeV; more comments about the precision of the results will be given later.

To complete the survey of possible Higgs masses, it is worth to make contact with the usual MSSM calculations with msugra or gauge-mediated boundary conditions imposed at a scale $\Lambda_S$ above the threshold $\Lambda$ of the heavy superpartners. Of course, to account for an inverted hierarchy in the scalar superpartners, the usual boundary conditions have to be modified. These calculations require to integrate the RG equations across the Effective SUSY-MSSM threshold and impose boundary conditions for the massive parameters at the scale $\Lambda_S$. The dimensionless couplings can be integrated up till the scale $\Lambda$ as explained in \S~\ref{sec:Unification}; at this scale, the supersymmetric boundary conditions fix all the dimensionless MSSM parameters, which can then be continued up to $\Lambda_S$. Regarding the massive parameters, the computation is done by modifying  as follows the iterated procedure described earlier yielding the correct electroweak symmetry breaking. It was mentioned earlier 
that electroweak symmetry breaking in a single Higgs decoupling limit fixes $m^2_H$ in the Effective SUSY theory. This, in turn, demanding that heavy MSSM Higgs states have masses equal to $\Lambda$, determines $m^2_{H_u}, m^2_{H_d}$ and $B_\mu$ in terms of $m^2_H,\alpha,\Lambda$, as explained in \S~\ref{subapp:higgsdecoupling} --see eq.~\eqref{eq:Higgsmatching}. The calculation starts by assigning guess values at the scale $\Lambda$ to all massive MSSM parameters. These are then matched with Effective SUSY parameters using the results of \S~\ref{subapp:matchhigh}, which are evolved downwards until reaching the lowest Effective SUSY threshold $\Lambda_{SM}$. At this scale, electroweak symmetry breaking is imposed as explained earlier, which determines a new value of $m^2_H(\Lambda_{SM})$ . All parameters are evolved upwards again, matched across the threshold at $\Lambda$ and evolved with the MSSM RG equations until reaching the scale $\Lambda_S$, where the desired boundary conditions are imposed. All 
parameters are then iteratively evolved up and down between the scales $\Lambda_{SM}$ and $\Lambda_S$, performing the correct matching at the intermediate scale $\Lambda$, until convergence is achieved between the results of successive determinations of $m^2_H$ at the scale $\Lambda_{SM}$.

\begin{figure}[h!]\centering
\begin{minipage}{.5\textwidth}\centering
\includegraphics[scale=.9]{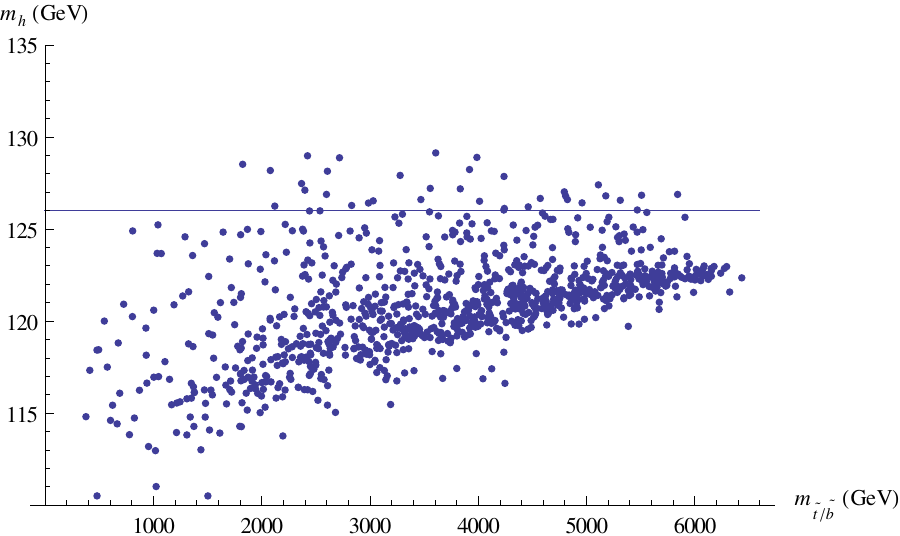}
\end{minipage}\begin{minipage}{.5\textwidth}\centering
   \includegraphics[scale=.9]{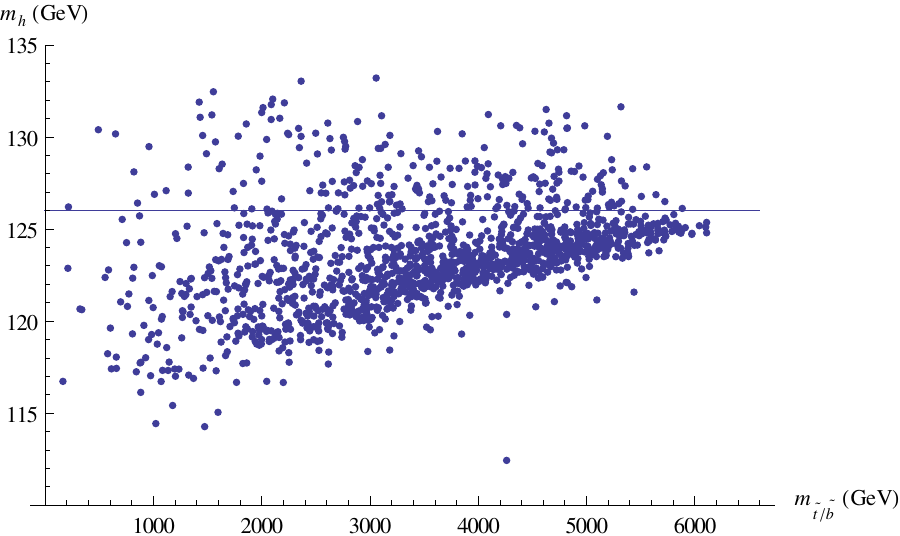}
\end{minipage}
\caption{\label{fig:scanmsugra} Higgs masses versus  the mass of the lightest colored scalar in minimal (left), and nonminimal (right) Effective SUSY scenarios with the boundary conditions of eq.~\eqref{eq:msugrabc}.}
\end{figure}

The following msugra-inspired boundary conditions, modified to make connection with  Effective SUSY scenarios, were considered:

\begin{align}
\label{eq:msugrabc}
\begin{array}{c}
\text{ minimal Effective SUSY}\\
\mu=M_1=M_2=M_3=m_F,\\
 { m^2_{q/u/d/l/e}}_{11}= { m^2_{q/u/d/l/e}}_{22}= {m^2_{d/l/e}}_{33}=\Lambda^2,\\
{ m^2_{q/u}}_{33}=m_s^2,\\
\frac{a_u}{y_t}=\frac{a_d}{y_b}=\frac{a_l}{y_\tau}=a_0,
\end{array}\quad\begin{array}{c}
\text{ nonminimal Effective SUSY}\\
\mu=M_1=M_2=M_3=m_F,\\
 { m^2_{q/u/d/l/e}}_{11}= { m^2_{q/u/d/l/e}}_{22}=\Lambda^2,\\
{ m^2_{q/u/d/l/e}}_{33}=m_s^2,\\
\frac{a_u}{y_t}=\frac{a_d}{y_b}=\frac{a_l}{y_\tau}=a_0,
\end{array}
\end{align}
with the trilinear couplings $a_u,a_d,a_l$ referring exclusively to the third generation. As explained earlier, $m^2_{H_u}, m^2_{H_d}$ and $B_\mu$ are not fixed at the high scale but obtained from the rest of parameters by demanding a correct electroweak symmetry breaking and Higgs decoupling limit. Fig.~\ref{fig:scanmsugra} shows the resulting Higgs masses after a scan over  $|m_F|$ between 200 and 1500 GeV, $m_s$ between 3500 and 10000 GeV (large values of $m_s$ are needed to avoid tachyons, see \S~\ref{sec:Tachyons}), $|a_0|$ between 0 and 20 TeV (to allow $|a_0(q\sim15TeV)|\lesssim4500$ GeV, as in the previous scans in the low energy Effective SUSY theories),  $\Lambda$ between 10 at 20 TeV, and $\cot\alpha\sim\tan\beta$ taking values between 5 and 60.

\begin{figure}[h!]\centering
\begin{minipage}{.5\textwidth}\centering
\includegraphics[scale=.9]{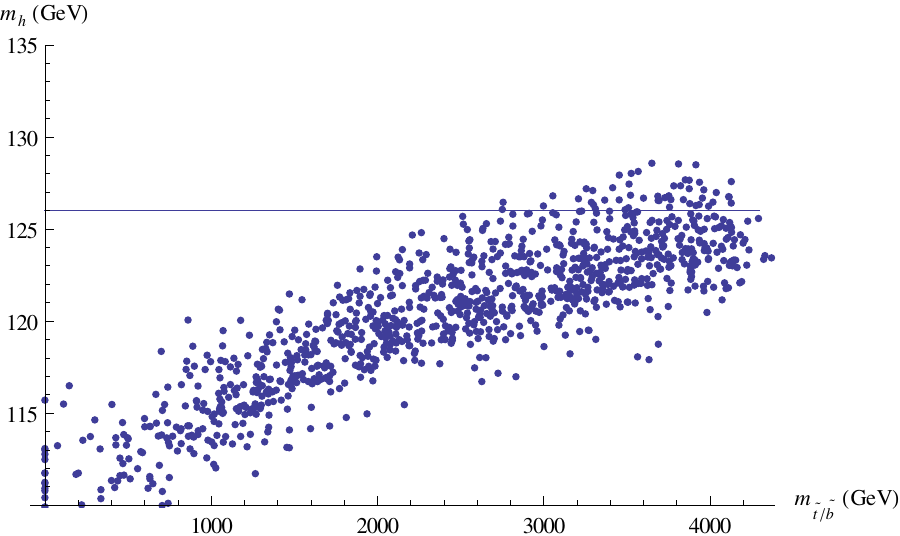}
\end{minipage}\begin{minipage}{.5\textwidth}\centering
   \includegraphics[scale=.9]{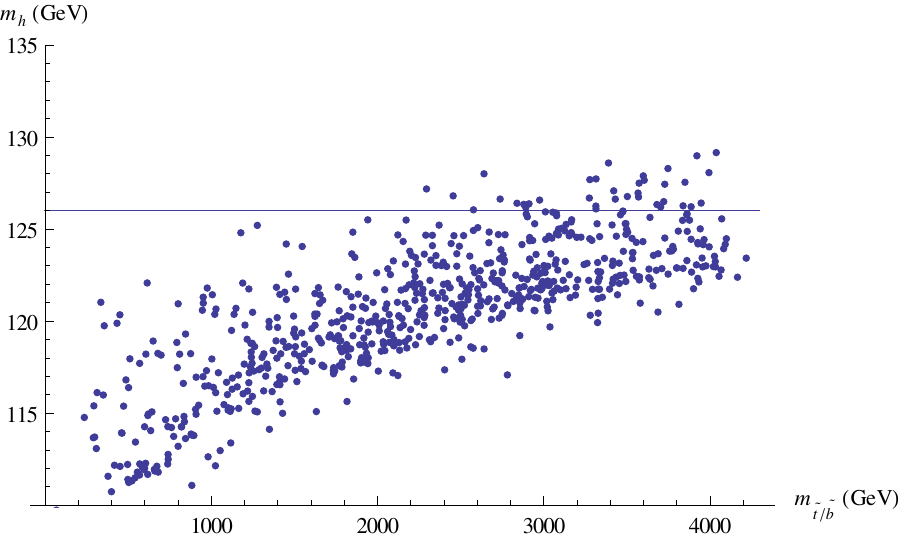}
\end{minipage}
\caption{\label{fig:scangm} Higgs masses versus  the mass of the lightest colored scalar in minimal (left), and nonminimal (right) Effective SUSY scenarios with the boundary conditions of eq.~\eqref{eq:gmbc}.}
\end{figure}

Boundary conditions inspired by gauge-mediation (at least for the light scalar and fermion masses) were also considered,
\begin{align}
\label{eq:gmbc}\begin{array}{c}
 \text{minimal Effective SUSY}\\
M_i=g^2_i\Lambda_g,\\
{ m^2_{q/u/d/l/e}}_{11}={ m^2_{q/u/d/l/e}}_{22}={ m^2_{d/l/e}}_{33}=\frac{\Lambda_S^2}{16\pi^2},\\
{ m^2_{i}}_{33}= \Lambda^2_G\sum_k g_k^4 C^k_2(i),\,i=q,u,\\
a_u=a_d=a_l=0,
\end{array}\quad\begin{array}{c}
 \text{nonminimal Effective SUSY}\\
M_i=g^2_i\Lambda_g,\\
{ m^2_{q/u/d/l/e}}_{11}={ m^2_{q/u/d/l/e}}_{22}=\frac{\Lambda_S^2}{16\pi^2},\\
 { m^2_{i}}_{33}= \Lambda^2_G\sum_k g_k^4 C^k_2(i),\\
a_u=a_d=a_l=0,
\end{array}
\end{align}
where   $\Lambda_S$ can be thought of as the SUSY breaking scale, and $\mu$  is taken as an extra parameter. Independent scales for gaugino and light squark masses were considered since the scale of R-symmetry breaking determining gaugino masses can be different from the scale determining squark masses in gauge-mediated models. A scan was performed over $\Lambda_S$ between 125 and 250 TeV, $|\Lambda_g|$ between 200 and 1600 GeV (taking $\mu=M_1$ at the high scale), $\Lambda_G$ between 500 and 4500 GeV and $\cot\alpha\sim\tan\beta$ between 5 and 60, with  Fig.~\ref{fig:scangm} shows the resulting Higgs masses plotted against the mass of the lightest colored state.

\begin{figure}[h!]\centering
 \begin{minipage}{.5\textwidth}\centering
   \includegraphics[scale=.85]{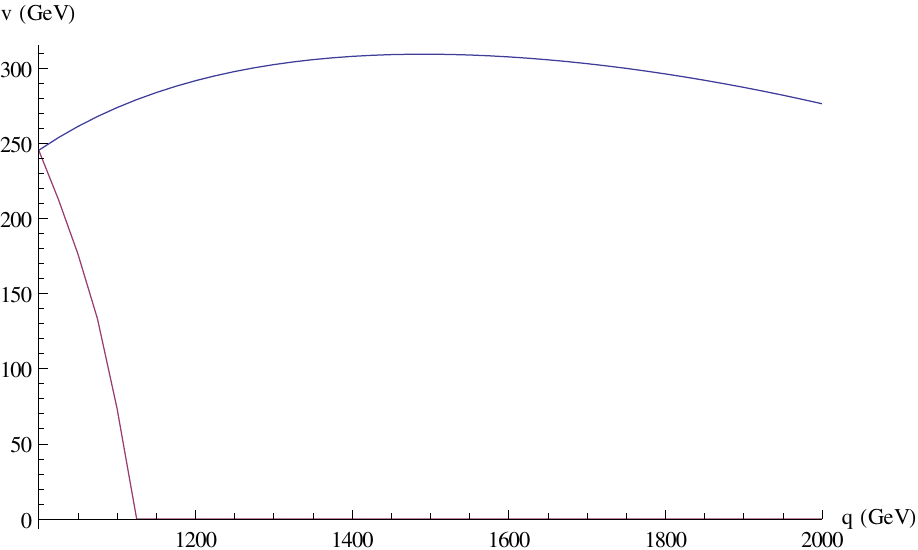}
\end{minipage}\begin{minipage}{0.5\textwidth}\centering
 \includegraphics[scale=.85]{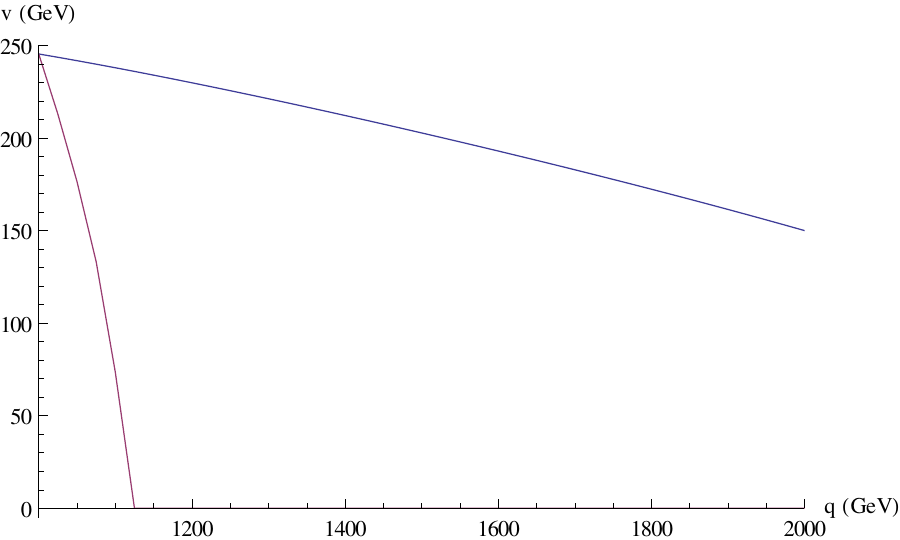}
\end{minipage}
\caption{\label{fig:Veffscale} Scale dependence of the Higgs VEV computed in a minimal Effective SUSY scenario (left) and a nonminimal one (right). The upper blue lines represent the results inlcuding two-loop effects due to scalars, while the lower purple lines were obtained with just the one-loop potential. In all cases it was required that the correct Higgs VEV was to be obtained at 1000 GeV.}
\end{figure}

In order to check the consistency of the results, it is possible to study the dependence with scale of the Higgs VEV obtained by minimizing the effective potential. Although in the calculations the Higgs VEV was fixed at its experimentally derived value when minimizing the potential at the lower Effective SUSY threshold --the minimization was used to solve for $m^2_H$ instead of $\langle H\rangle$-- one may still use the resulting RG flow to evaluate the effective potential at another scale, keeping $\langle H\rangle$ arbitrary and finding the value corresponding to the minimum at that scale. Since the full effective potential is scale-independent, by using higher order corrections in the effective potential, the scale dependence of  $\langle H\rangle$ should improve. Fig.~\ref{fig:Veffscale}  shows that this is  the case indeed: the two-loop scalar corrections included in the calculations of the effective potential do improve the scale dependence of $\langle H\rangle$ with respect to the one-loop result. 

Finally, in order to estimate the precision of the calculation, the effect of changing the value of $\alpha_s$ within its error or the difference between the one-loop value and the one inlcuding 2 loop corrections due to scalars is in general less than half a GeV. The one-loop and two-loop result tend to differ more significantly for large trilinear couplings and very different values of $m^2_Q$ and $m^2_U,m^2_D$, giving rise to light colored stops/sbottoms; in this case  discrepancies can reach 3 GeV for lightest sbottoms around 300-400 GeV. It should also be noted that the calculations ignored thresholds corrections other than those due to a change of renormalization scheme at the scale $\Lambda$ at which the effective theories are matched with the MSSM.  The most important threshold contributions for the Higgs mass will be those of $\lambda$, since the quartic coupling fixes the tree-level value of the mass. These threshold contributions can be obtained from the contributions of the scalars to the one-
loop 
effective potential for the MSSM,  upon substituting $H_u\sim \cos\alpha H, \,H_d\sim \sin\alpha H^\dagger$. Neglecting the Higgs  dependence of the mass matrices of the heavy charged and neutral Higgs fields (or choosing a scale to evaluate the effective potential in which their contribution vanishes) and ignoring the first and second generation Yukawas and trilinear couplings, one gets:
\begin{align*}
\delta\lambda_{nmES}=&\frac{1}{3200\pi ^2}\cos2\alpha^2 \left(6 g_1^4 \log\frac{{m^2_d}_1}{q^2}+6 g_1^4 \log\frac{{m^2_d}_2}{q^2}+18 g_1^4 \log\frac{{m^2_e}_1}{q^2}+18 g_1^4 \log\frac{{m^2_e}_2}{q^2}\right.\\
&+9 g_1^4 \log\frac{{m^2_l}_1}{q^2}+25 g_2^4 \log\frac{{m^2_l}_1}{q^2}+9 g_1^4 \log\frac{{m^2_l}_2}{q^2}+25 g_2^4 \log\frac{{m^2_l}_2}{q^2}+3 g_1^4 \log\frac{{m^2_q}_1}{q^2}\\
&\left.+75 g_2^4 \log\frac{{m^2_q}_1}{q^2}+3 g_1^4 \log\frac{{m^2_q}_2}{q^2}+75 g_2^4 \log\frac{{m^2_q}_2}{q^2}+24 g_1^4 \log\frac{{m^2_u}_1}{q^2}+24 g_1^4 \log\frac{{m^2_u}_2}{q^2}\right)
\end{align*}
for nonminimal Effective SUSY scenarios and 
\begin{align*}
\delta\lambda_{mES}=&\delta\lambda_{nmES}+\frac{1}{6400 \pi ^2}\left\{12 \left(5 y_b^2+\left(g_1^2-5 y_b^2\right) \cos2\alpha\right)^2 \log\frac{{m^2_d}_3}{q^2}\right.\\
&+4 \left(5 y_\tau^2+\left(3 g_1^2-5 y_\tau^2\right) \cos2\alpha\right)^2 \log\frac{{m^2_e}_3}{q^2}+\left(9 g_1^4+30 g_1^2 y_\tau^2+25 \left(g_2^4-2 g_2^2 y_\tau^2+6 y_\tau^4\right)\right.\\
&-20 y_\tau^2 \left(3 g_1^2-5 g_2^2+10 y_\tau^2\right) \cos2 \alpha+\left(9 g_1^4+30 g_1^2 y_\tau^2+25 \left(g_2^4-2 g_2^2 y_\tau^2\right.\right.\\
&\left.\left.\left.\left.+2 y_\tau^4\right)\right) \cos4 \alpha\right) \log\frac{{m^2_l}_3}{q^2}\right\},
\end{align*}
for minimal ones. 
These corrections vanish in the limit in which the heavy soft masses are degenerate. In order to estimate the size of $\delta\lambda$, one may use the boundary conditions of eqs.~\eqref{eq:msugrabc} and \eqref{eq:gmbc} and run down the soft masses till the threshold $\Lambda$ of the heavy first and second generation sparticles. For example, in nonminimal scenarios, using the boundary conditions \eqref{eq:msugrabc} with $\cot\alpha=10$, $\Lambda=15$ TeV,   $m_s=6$ TeV,  $m_F=1$ TeV, $a_0=0$ at $10^{16}$ GeV, the result is $\delta\lambda_{nmES}=-7\cdot10^{-6}$, and using \eqref{eq:gmbc} with $\cot\alpha=10$, $\Lambda_S=200$ TeV, $\Lambda_G=3$ TeV, $\mu=\Lambda_g=1$ TeV, one gets $\delta\lambda_{nmES}=-1\cdot10^{-6}$,  corresponding to  changes in the Higgs mass squared of $\delta m^2=2\delta\lambda v^2\sim-0.4\, {\rm GeV}^2$ and $-0.06\, {\rm GeV}^2$, respectively, which for a Higgs mass around 120 GeV imply shifts of -0.002 GeV or smaller. The same boundary conditions in minimal scenarios imply shifts in the 
Higgs mass of 0.001 GeV or smaller.

\section{Fine-tuning estimates\label{sec:FineTuning}}

A possible measure of fine-tuning ${\Delta}$ in a theory with a single Higgs field is the following,
\begin{align}\label{eq:FTmeasure}
 {\Delta}=\frac{1}{\langle H \rangle^2}\sqrt{\sum_i \left|x_i\frac{\partial \langle H \rangle^2(x_i)}{\partial x_i}\right|^2},
\end{align}
where $x_i$ designate the values of the dimensionful parameters in the theory at a high energy boundary, which for Effective SUSY scenarios will be taken as the scale $\Lambda$ of the decoupled sparticles.

\begin{figure}[h!]\centering
\includegraphics[scale=.85]{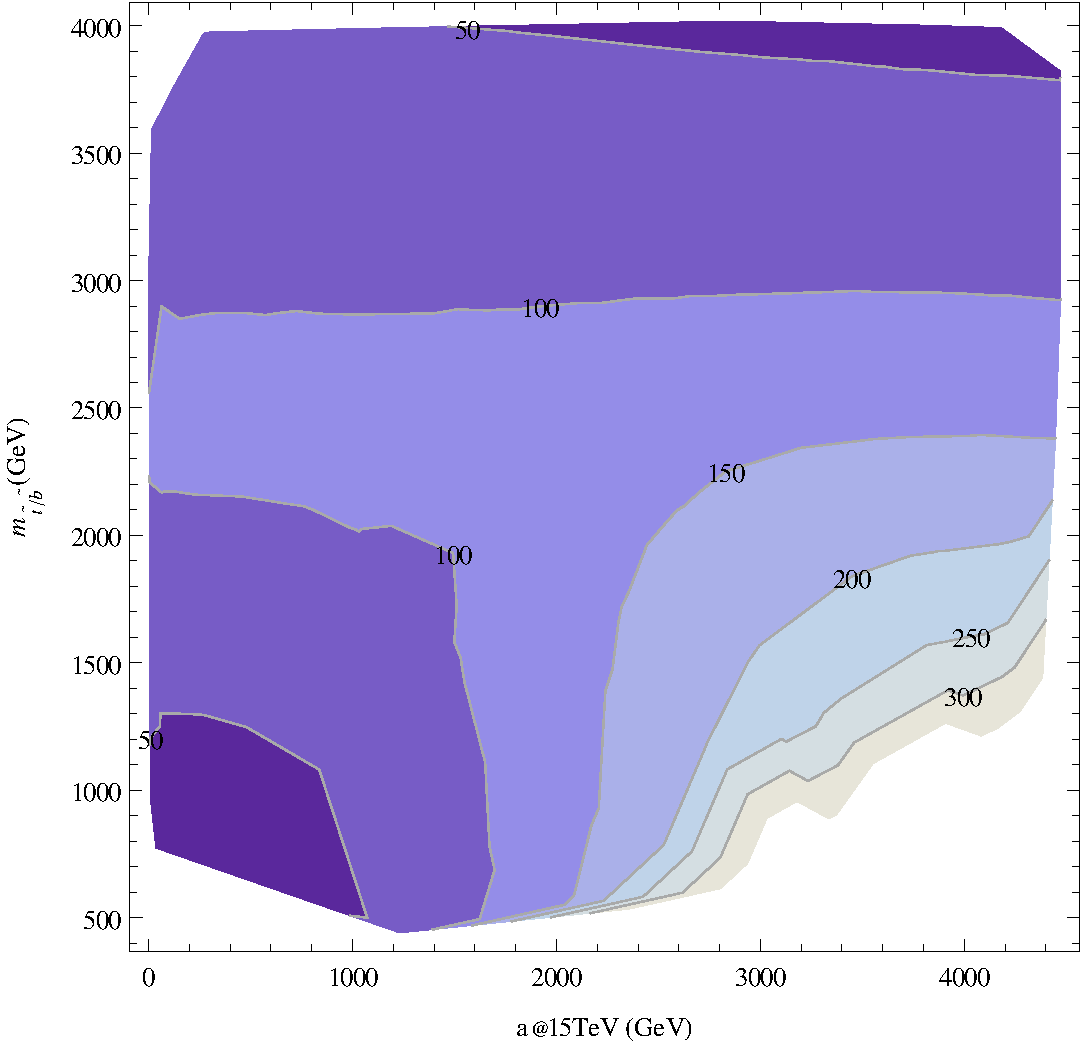}
\caption{\label{fig:FT} Fine-tuning estimations in nonminimal Effective SUSY scenarios in terms of the high scale boundary value of $a_u$, with $c_d=c_l=0$ at the high scale and the lightest stop/sbottom eigenvalue. Fermion mass parameters were fixed at 500 GeV in the boundary; common boundary values for the scalar soft masses were assumed.}
\end{figure}

In order to estimate fine-tuning in Effective SUSY scenarios, it was chosen to compute $\langle H \rangle$ from a simplified version of the one-loop effective potential, obtained by doing a series expansion in the Higgs field up to terms of mass dimension four, and ignoring the contributions of the fermion masses. The derivatives of $\langle H \rangle$ with respect to the high scale mass parameters were approximated as follows: in each case, the low scale dimensionful parameters resulting from the RG flow were fitted to linear combinations of the boundary values (or their products and powers, as dictated by the corresponding mass dimensions). This produced approximate analytical expressions for  $\langle H \rangle$ in terms of the high scale parameters that were used to compute ${\Delta}$. 
Fig.~\ref{fig:FT} shows the results for the fine-tuning in associated with the nonminimal Effective SUSY scenarios that yield the Higgs masses of Fig.~\ref{fig:scan2dES}.
The value of $\Lambda$ within these approximations is not very sensitive to $\cot\alpha$, and the results for minimal scenarios are very similar. It can be concluded thus that effective SUSY scenarios with a 126 GeV Higgs and trilinear couplings of the order of 3 TeV have a fine-tuning of the order of one part in 200, if the measure of eq.~\eqref{eq:FTmeasure} is used.

\section{Revisiting tachyon bounds for third-generation scalars \label{sec:Tachyons}}


As explained in the introduction, heavy first and second generation scalars tend to drive the running soft masses of the third generation towards negative values, which gives rise to lower bounds on the third generation masses at the susy breaking scale. This in turn implies a lower bound in the amount of fine-tuning of phenomenologically acceptable theories. Due to the large discrepancies between the RG flows of the MSSM in the $\overline{\rm DR}$ scheme and the flows in the Effective theories obtained by decoupling the heavy particles, these bounds have to be revisited. To do so, the 2 loop RG equations were run between a SUSY breaking scale $\Lambda_S$ and the lower Effective SUSY threshold $\Lambda_{SM}$, performing the matching across the intermediate threshold $\Lambda$ as explained in previous sections.  Both the cases of high and low scale SUSY breaking were considered. 

\begin{figure}[h!]\centering
\begin{minipage}{.5\textwidth}\centering
   \includegraphics[scale=.89]{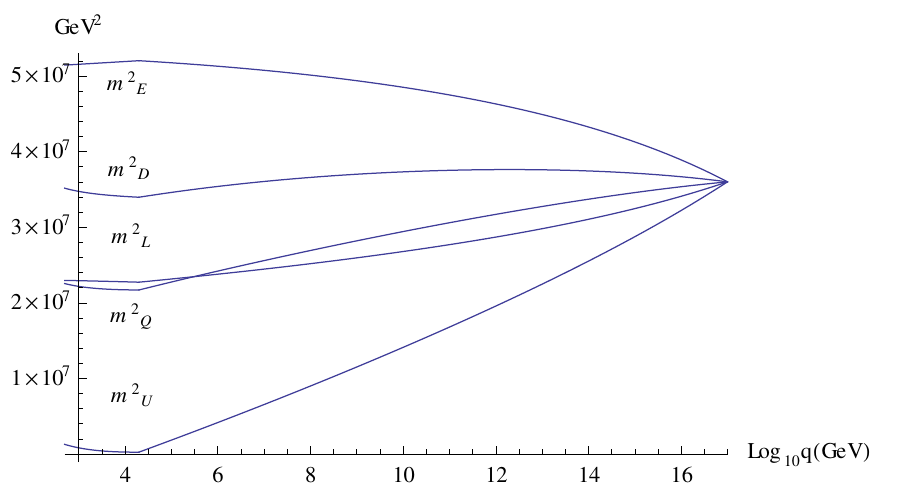}
\end{minipage}\begin{minipage}{0.5\textwidth}\centering
   \includegraphics[scale=.89]{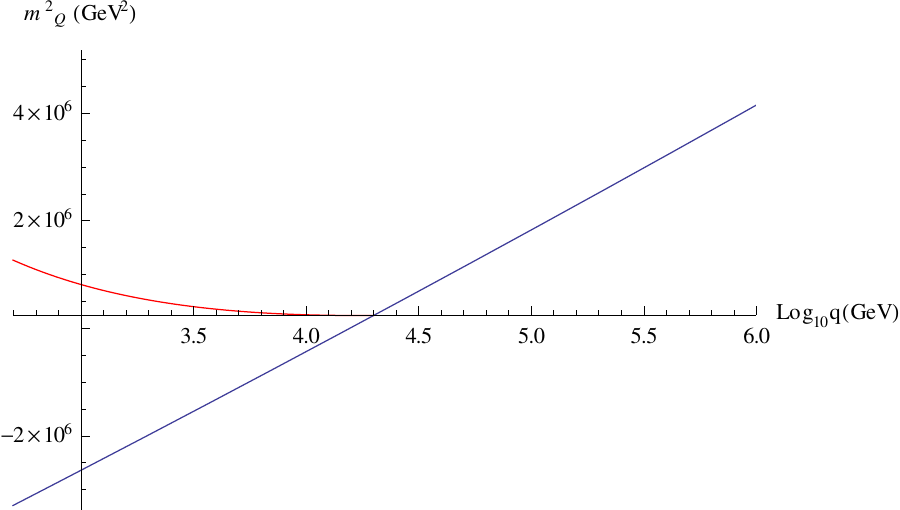}
\end{minipage}
\caption{\label{fig:ESRGconst} RG evolution for the soft masses in a nonminimal Effective SUSY scenario with the boundary conditions of eq.~\eqref{eq:msugrabc}. On the right, the blue line represents the MSSM running mass, which becomes tachyonic, while the red line corresponds to the  running mass in the theory with the heavy particles decoupled.}
\end{figure}

\begin{figure}[h!]\centering
 \begin{minipage}{.5\textwidth}\centering
   \includegraphics[scale=.85]{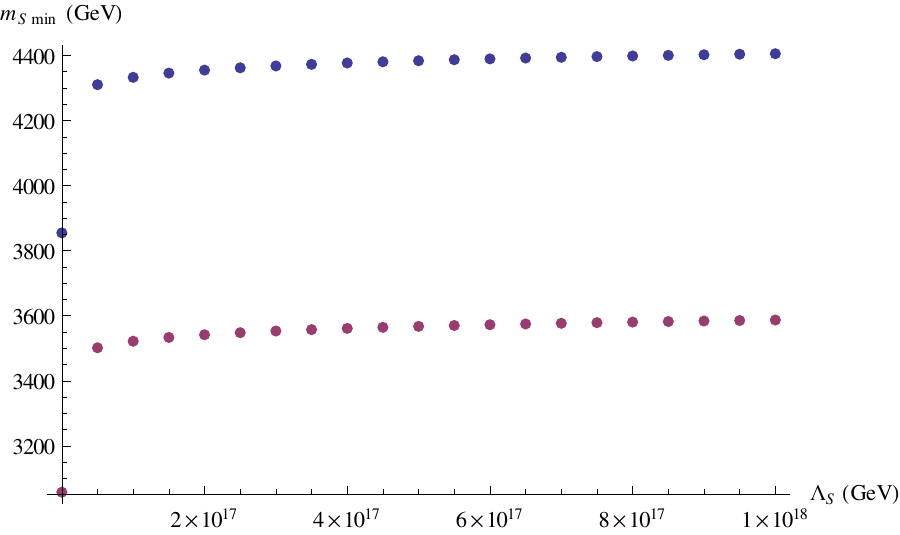}
\end{minipage}\begin{minipage}{0.5\textwidth}\centering
 \includegraphics[scale=.85]{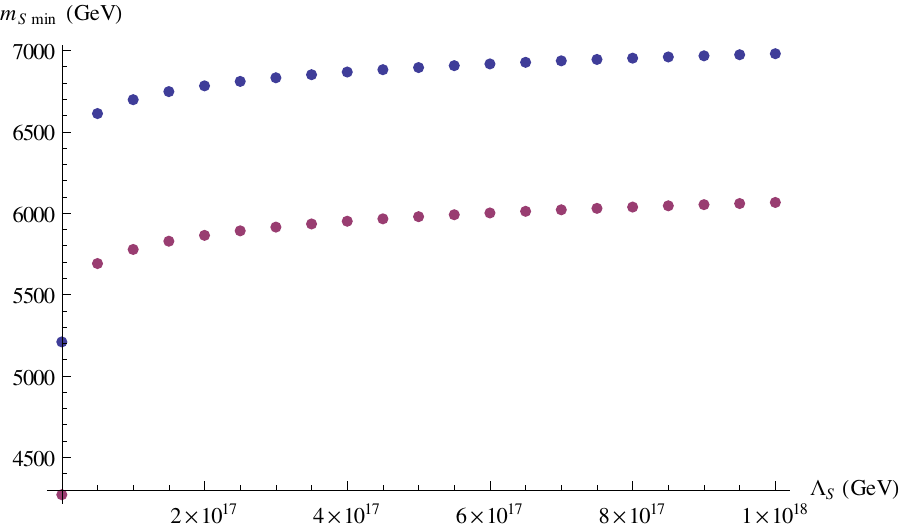}
\end{minipage}
\caption{\label{fig:msugrabounds} Minimum value of the scalar mass $m_s$ needed to avoid tachyonic soft masses at 500 GeV in terms of the high scale $\Lambda_S$, in minimal (left) and nonminimal (right) Effective SUSY scenarios with the boundary conditions of eq.~\eqref{eq:msugrabc}. The upper blue dotes correspond the the MSSM RG flow, and the lower ones to the flow implementing the decoupling the heavy particles.}
\end{figure}

In the case of high scale breaking, the boundary conditions of eq.~\eqref{eq:msugrabc} were considered. It is known that in such case the RG drives $m^2_{q_3}$ to tachyonic values, which forces a lower bound in the high scale mass $m_s$ of the light scalars. Fig.~\ref{fig:ESRGconst} shows the running soft masses in a nonminimal Effective SUSY scenario.  Fig.~\ref{fig:msugrabounds} shows the bounds obtained in minimal and nonminimal Effective SUSY scenarios as a function of the high scale $\Lambda_S$ versus the ones obtained with the MSSM RG equations. $m_F$ was taken as 1 TeV, $\Lambda$ as 20 TeV and $a_0=0$. As anticipated, there are large differences of up to 900 GeV.

\begin{figure}[h!]\centering 
\begin{minipage}{.5\textwidth}\centering
   \includegraphics[scale=.9]{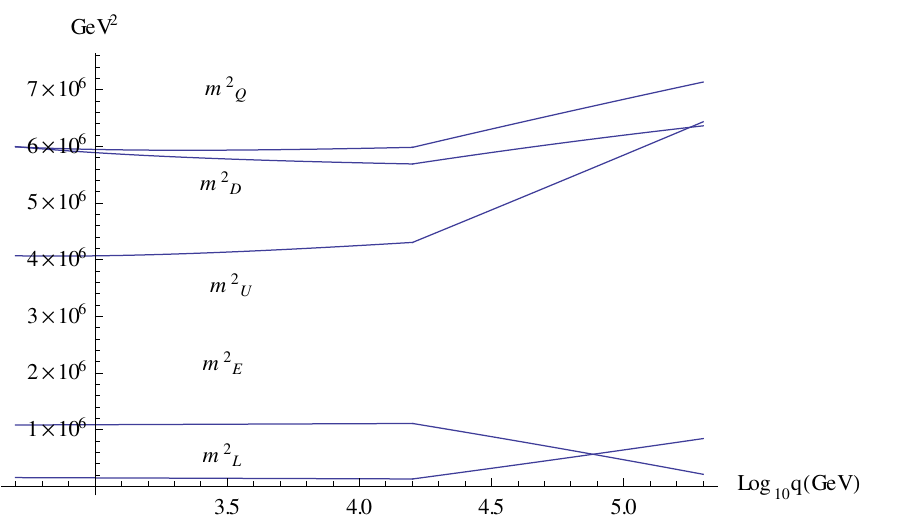}
\end{minipage}\begin{minipage}{0.5\textwidth}\centering
   \includegraphics[scale=.9]{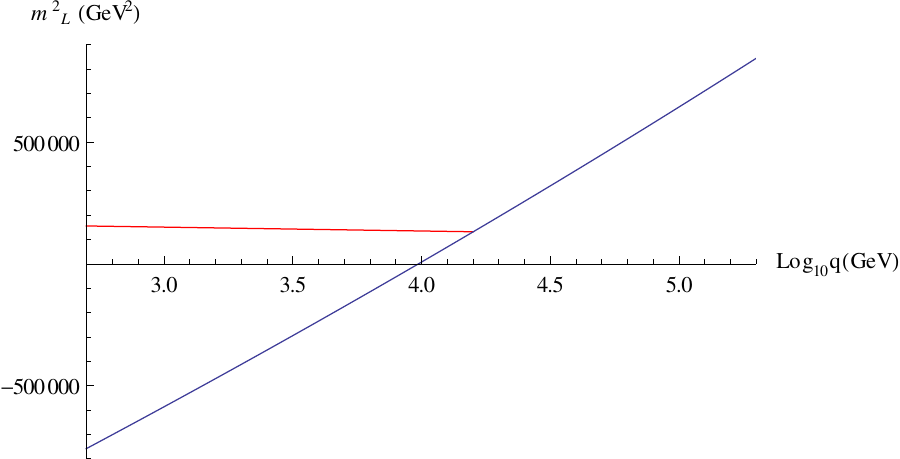}
\end{minipage}
\caption{\label{fig:ESRGGM} RG evolution for the soft masses in a nonminimal Effective SUSY scenario with the boundary conditions of eq.~\eqref{eq:gmbc}. On the right, the blue line represents the MSSM running mass, which becomes tachyonic, while the red line corresponds to the  running mass in the theory with the heavy particles decoupled.}
\end{figure}

\begin{figure}[h!]\centering
 \begin{minipage}{.5\textwidth}\centering
   \includegraphics[scale=.85]{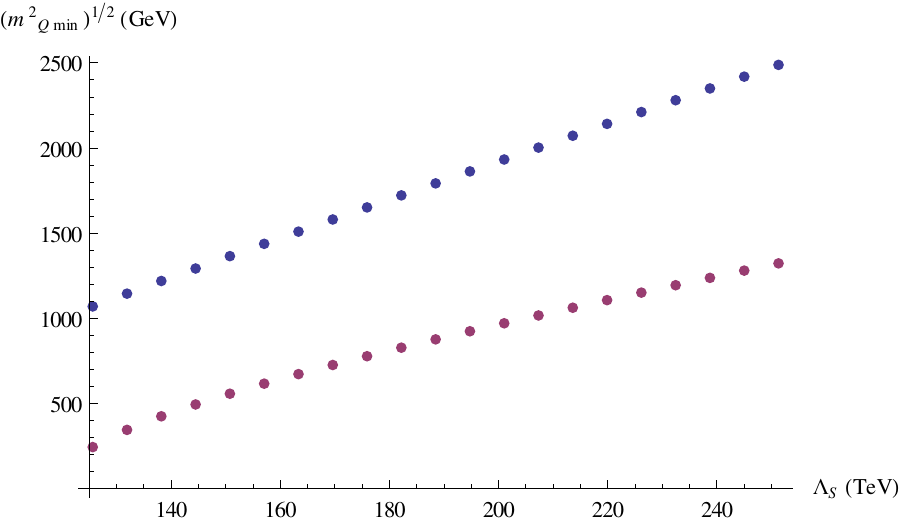}
\end{minipage}\begin{minipage}{0.5\textwidth}\centering
 \includegraphics[scale=.85]{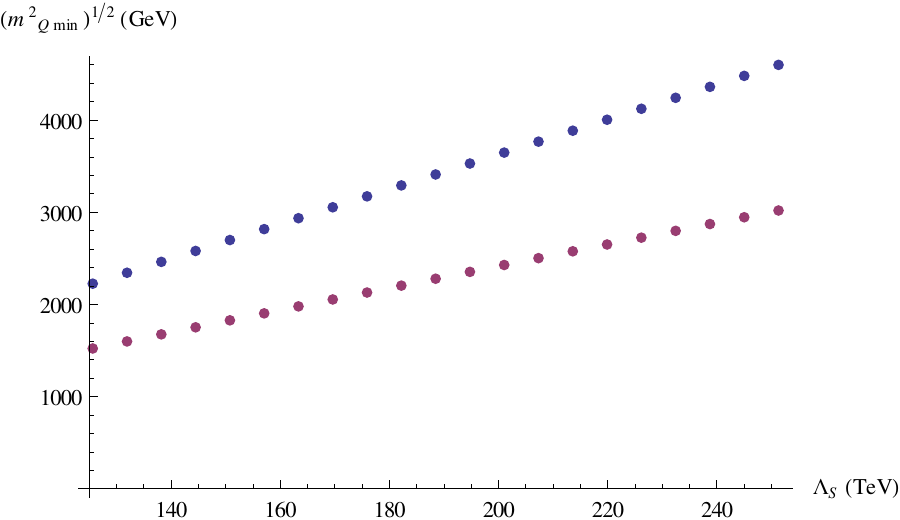}
\end{minipage}
\caption{\label{fig:GMbounds} Left: Minimum value of the scalar mass $m^2_(\Lambda_S)$ needed to avoid tachyonic soft masses at 500 GeV in terms of the high scale $\Lambda_S$, in minimal (left) and nonminimal (right) Effective SUSY scenarios with the boundary conditions of eq.~\eqref{eq:gmbc}. The upper blue dotes correspond the the MSSM RG flow, and the lower ones to the flow implementing the decoupling the heavy particles.}
\end{figure}

As pertains to low scale susy breaking, similar calculations were done using the boundary conditions of eq.~\eqref{eq:gmbc}. In this case, the RG flow tends to drive $m^2_U$ tachyonic in minimal Effective SUSY scenarios, while in nonminimal ones it is $m^2_L$ that tends to negative values, as can be seen in Fig.~\ref{fig:ESRGGM}. The minimal value of $m^2_Q(\Lambda_S)$ for which there are no negative soft masses at the scale $\Lambda_{SM}$ (chosen as 500 GeV) is plotted against the scale  $\Lambda_S$ in  Fig.~\ref{fig:GMbounds} and compared with the bound obtained with the MSSM RG equations. $\Lambda_g$ was taken as 1 TeV. As before, the difference between the bounds obtained with the MSSM ${\overline{\rm DR}}$ RG equations and the decoupled flow is significant, between 600 and 1600 GeV.

\newpage
\section{Conclusions \label{sec:conclusions}}

This paper has examined some basic phenomenological aspects of Effective SUSY scenarios with a single Higgs field at low energies by using an approach that takes into account the decoupling of heavy sparticles. Two types of Effective SUSY scenarios were considered: the minimal one, in which the only light third generation scalars are the two stop states and the left-handed sbottom, and a nonminimal scenario in which all third generation scalars are light. The two-loop renormalization group equations obtained in ref.~\cite{Tamarit:2012} were used below the scales of the heavy sparticles, and the well known MSSM $\overline{\rm DR}$ RG equations \cite{Martin:1993zk} above it; various one-loop threshold effects were included in the analysis. The aspects that were studied are: gauge coupling and $b$-$\tau$ unification, Higgs masses, fine-tuning and tachyon bounds. The results can be summarized as follows:

Gauge coupling unification at two loops in Effective SUSY scenarios is not spoiled by the large hierarchies in the scalar masses; it typically holds within a precision better than 2\%, which is  similar to or better than the  3\% or less accuracy corresponding to standard MSSM scenarios --see fig.~\ref{fig:unifscansall}. Most of the Effective SUSY realizations yield a negative value for the GUT scale threshold parameter $\epsilon_g$ of eq.~\eqref{eq:epsilong}, though in nonminimal scenarios large trilinear couplings $c_d,c_l$ can give rise to positive values of $\epsilon_g$ resulting from large sbottom mixing angles affecting the threshold corrections to the bottom Yukawa coupling.

Concerning $b$-$\tau$ unification, it seems hard to achieve in minimal Effective SUSY scenarios, where threshold corrections of around 20\% would be needed; the Yukawa couplings tend to meet at scales far below the gauge coupling unification scale. A different situation holds in nonminimal Effective SUSY scenarios; again, the effect of the trilinear couplings $c_d,c_l$ is important, and nonzero sbottom mixing angles allow the $y_b$ and $y_\tau$ too meet at the gauge coupling unification scale (see Fig.~\ref{fig:btau}).
 
Higgs masses were also computed using the two-loop Renormalization group equations together with the full one-loop effective potential supplemented with the two-loop contributions due to scalars. A 125 GeV Higgs with third generation squarks below 5 TeV requires in general large trilinear couplings, of the order of 2 TeV (see figs.~\ref{fig:scansMES} and \ref{fig:scansES}). Nonminimal scenarios allow for a wider spread in Higgs masses, which signals that the effects of couplings in the down sector (particularly trilinear couplings) can become important. 
In order to have a 125 GeV Higgs correlated with light colored scalars (at around 300 GeV or even below), non-universal boundary conditions for the soft masses and trilinear couplings  at the scale of the heavy sparticles are preferred. Imposing boundary conditions inspired by msugra or gauge mediation (but adapted to Effective SUSY scenarios) at a scale higher than that of the heavy sparticles, getting a 125 GeV Higgs 
requires in general heavier stops/sbottoms (see figs.~\ref{fig:scanmsugra} and \ref{fig:scangm}), particularly with gauge mediated third generation soft masses and zero a-terms, for which the lightest third generation scalar has to be above $\sim2$ TeV in nonminimal Effective SUSY scenarios and above $\sim3$ TeV in minimal ones --this bound is, however, much lower than in traditional gauge mediated scenarios, for which it lies around 10 TeV \cite{Draper:2011aa}.

Fine-tuning in Effective SUSY scenarios was also estimated with respect to the boundary conditions at the scale of the heavy sparticles, and shown to be of the order of one part in two or three hundred in regions of parameter space with large $a_u$ and light stop/sbottoms, if the measure of eq.~\eqref{eq:FTmeasure} is used.

Finally, it was shown that there are significant discrepancies between the MSSM $\overline{\rm DR}$ running of the soft masses and their renormalization group evolution under the beta functions implementing decoupling (see figs.~\ref{fig:ESRGconst} and \ref{fig:ESRGGM}). In particular, tachyon bounds caused by the large negative contributions of the heavy scalars to the evolution of the third generation soft masses towards low scales can be relaxed by up to 900 GeV in models with high scale SUSY breaking with the boundary conditions of eq.~\eqref{eq:msugrabc} and up to 1.5 TeV or more in models of low scale SUSY breaking with the boundary conditions of eq.~\eqref{eq:gmbc} (see figs.~\ref{fig:msugrabounds} and \ref{fig:GMbounds}).
\section*{Acknowledgements}
The author wishes to thank Sakura Sch\"afer-Nameki, Joel Jones, Gianluca Blankenburg and the members of the Particle Physics group at Perimeter Institute for useful conversations. Research
at the Perimeter Institute is supported in part by the Government of Canada through
NSERC and by the Province of Ontario through MEDT. This work was financed in part by the Spanish Ministry of Science and Innovation through project FPA2011-24568.

\newpage
\appendix

\section{High scale matching \label{app:highscale}}

In this section, the ${\overline{\rm MS}}$ couplings of the low energy effective theories are matched at the scale of the heavy heavy scalars with the  MSSM couplings in the SUSY-preserving ${\overline{\rm DR}}$ scheme, in the decoupling limit with a single Standard Model-like Higgs. One-loop factors due to the change of scheme are included, and they are obtained from the formulae of ref.~\cite{Martin:1993yx}. Additional one-loop thresholds for the gauge couplings due to integrating out fields across the Effective SUSY-MSSM threshold are also taken into account, following ref.~\cite{Hall:1980kf}.

\subsection{MSSM Higgs decoupling limit\label{subapp:higgsdecoupling}}
The Effective SUSY theories considered in this paper have a single, Standard-Model like Higgs field at low energies, and therefore have to be matched with the MSSM in a Higgs decoupling limit. Consider integrating out a heavy Higgs field in the MSSM. The quadratic pieces in the MSSM Higgs potential read
\begin{align*}
 V_{MSSM}\supset (\mu^2+m^2_{H_u})|H_u|^2+(\mu^2+m^2_{H_d})|H_d|^2-(B_\mu H_u H_d+{\rm c.c.}).
\end{align*}
Ignoring Higgs VEVs and phases, this yields a mass matrix with eigenvalues
\begin{align*}
 m_{H/{\mathcal H}}=\frac{1}{2}\left(m^2_{H_u}+m^2_{H_d}+2\mu^2\pm\sqrt{4 B_\mu^2+(m^2_{H_d}-m^2_{H_u})^2}\right),
\end{align*}
and a mixing angle $\alpha$
\begin{align}
 \left.\begin{array}{c}
  H=\cos\alpha H_u+\sin\alpha H_d^\dagger,\\
{\mathcal H}=\sin\alpha H_u-\cos\alpha H_d^\dagger,
 \end{array}\right\}\quad\tan\alpha=\frac{2B_\mu}{m^2_{H_d}-m^2_{H_u}+\sqrt{4B_\mu^2+(m^2_{H_d}-m^2_{H_u})^2}}.\label{eq:alpha}
\end{align}

The light field $H$ is identified with the Higgs field in the Effective SUSY scenarios considered in the paper. Imposing that the heavy eigenvalue is equal to the scale $\Lambda$ of the heavy scalars (the cutoff of the effective theories) one can get the following identities relating the MSSM parameters $B_\mu, m^2_{H_d}, m^2_{H_u}$ with the effective theory parameters $\Lambda, m^2_H, \alpha$ ($\alpha$ can be considered as a parameter in the low energy effective theory since it will enter the boundary conditions at the scale $\Lambda$, as will be seen in \S~\ref{subapp:matchhigh}):
\begin{align}
\nonumber m^2_{H_u}=&\frac{m^2_H+\Lambda^2\tan^2\alpha}{1+\tan^2\alpha},\\
\label{eq:Higgsmatching}m^2_{H_d}=&\Lambda^2\cos^2\alpha+m^2_H\sin^2\alpha,\\
\nonumber B_\mu=&(\Lambda^2-m^2_H)\cos\alpha\sin\alpha.
\end{align}
In the decoupling limit, the VEV of the very massive field $\cal H$ should be stabilized at zero. Using eq.~\eqref{eq:alpha} and demanding $\langle{\cal H}\rangle=0$, one can relate $\beta=\arctan\frac{\langle H_u\rangle}{\langle H_d \rangle}$ with $\alpha$:
\begin{align}
\sin(\alpha+\beta)=1, \label{eq:alphabeta}
\end{align}
which is precisely the condition in the MSSM that guarantees a Standard Model Higgs decoupling limit. This allows to match all the parameters in the Higgs sector across the Effective SUSY-MSSM threshold. The boundary conditions for the rest of the parameters are analyzed in the following sections.

\subsection{Boundary conditions - matching of couplings\label{subapp:matchhigh}}
Supersymmetry relates the scalar-gaugino-fermion couplings to the gauge couplings, and it also establishes a relationship between the quartic couplings and the gauge and Yukawa couplings. These relations should be imposed as boundary conditions at the cutoff scale of the decoupled Effective SUSY theories. There is yet a subtlety: the SUSY relations hold in a SUSY preserving renormalization scheme, such as $\overline{\rm DR}$, while the RG equations of the low-energy theories have been obtained in an $\overline{\rm MS}$ scheme. Therefore, the boundary conditions have to include one-loop conversion factors between both schemes.

In principle, one could consider additional thresholds for the gauge couplings coming from integrating out heavy fields \cite{Hall:1980kf}, but these vanish if all the heavy particles are degenerate and the threshold sits at the scale of their mass. They will thus be neglected at the MSSM matching scale, though they will be considered at the electroweak scale.

In order to compare the quartic couplings in the MSSM with those in the Lagrangians of eq.~\eqref{eq:LagrangianMES} and \eqref{eq:LagrangianNMES}, it is useful to use the expressions in ref.~\cite{Tamarit:2012} that specify how to absorb redundant couplings into the chosen set of independent couplings of the low energy Lagrangians. Here the final results will be provided. 
In the following formulae, the MSSM  $\overline{\rm DR}$ couplings will be denoted with tildes, and the Effective SUSY ones without them. The matching formulae below are understood to be evaluated at the scale $\Lambda$ of the heavy scalars. The boundary conditions neglect flavour-mixing effects.

\begin{align*}
 g_k=&\tilde g_k\left(1-\frac{g_k^2}{96\pi^2}C(G_k)\right),\quad k=1,2,3,\\
y_t=&\sin\beta\,\tilde y_t\left(1+\frac{1}{32\pi^2}\left(\frac{8}{3}g_3^2-\frac{3}{4}g_2^2-\frac{1}{60}g_1^2\right)\right),\\
y_b=&\cos\beta\,\tilde y_b\left(1+\frac{1}{32\pi^2}\left(\frac{8}{3}g_3^2-\frac{3}{4}g_2^2-\frac{13}{60}g_1^2\right)\right),\\
y_\tau=&\cos\beta\,\tilde y_\tau\left(1+\frac{1}{32\pi^2}\left(-\frac{3}{4}g_2^2+\frac{9}{20}g_1^2\right)\right),\\
z_{q_i}=&\delta_{i3}\,\tilde y_t\left(1+\frac{1}{32\pi^2}\left(-\frac{4}{3}g_3^2+\frac{3}{2}g_2^2-\frac{11}{30}g_1^2\right)\right),\\
z_{u_i}=&\delta_{i3}\,\tilde y_t\left(1+\frac{1}{32\pi^2}\left(-\frac{4}{3}g_3^2-\frac{3}{4}g_2^2+\frac{23}{60}g_1^2\right)\right),\end{align*}\begin{align*}
z_{d_i}=&\delta_{i3}\,\tilde y_b\left(1+\frac{1}{32\pi^2}\left(-\frac{4}{3}g_3^2-\frac{3}{4}g_2^2+\frac{11}{60}g_1^2\right)\right),\\
z_{q^*_i}=&\delta_{i3}\,\tilde y_b\left(1+\frac{1}{32\pi^2}\left(-\frac{4}{3}g_3^2+\frac{3}{2}g_2^2+\frac{1}{30}g_1^2\right)\right),\\
z_{l_i}=&\delta_{i3}\,\tilde y_\tau\left(1+\frac{1}{32\pi^2}\left(\frac{3}{2}g_2^2+\frac{3}{10}g_1^2\right)\right),\\
z_{e_i}=&\delta_{i3}\,\tilde y_\tau\left(1+\frac{1}{32\pi^2}\left(-\frac{3}{4}g_2^2+\frac{9}{20}g_1^2\right)\right),\\
g_{H_1}=&\sqrt{\frac{6}{5}}\cos\alpha\,\tilde g_1\left(1+\frac{1}{32\pi^2}\left(-\frac{3}{4}g_2^2-\frac{3}{20}g_1^2\right)\right),\\
g_{H_1^*}=&\sqrt{\frac{6}{5}}\sin\alpha\,\tilde g_1\left(1+\frac{1}{32\pi^2}\left(-\frac{3}{4}g_2^2-\frac{3}{20}g_1^2\right)\right),\\
g_{H_2}=&\sqrt{2}\cos\alpha\,\tilde g_2\left(1+\frac{1}{32\pi^2}\left(\frac{5}{4}g_2^2-\frac{3}{20}g_1^2\right)\right),\\
g_{H_2^*}=&\sqrt{2}\sin\alpha\,\tilde g_2\left(1+\frac{1}{32\pi^2}\left(\frac{5}{4}g_2^2-\frac{3}{20}g_1^2\right)\right),\\
g_{Q_{j,1}}=&\delta_{j3}\,\sqrt{\frac{6}{5}}\tilde g_1 \left(1+\frac{1}{32\pi^2}\left(-\frac{4}{3}g_3^2-\frac{3}{4}g_2^2-\frac{1}{60}g_1^2\right)\right),\\
g_{Q_{j,2}}=&\delta_{j3}\,\sqrt{2}\tilde g_2\left(1+\frac{1}{32\pi^2}\left(-\frac{4}{3}g_3^2+\frac{5}{4}g_2^2-\frac{1}{60}g_1^2\right)\right),\\
g_{Q_{j,3}}=&\delta_{j3}\,\sqrt{2}\tilde g_3\left(1+\frac{1}{32\pi^2}\left(\frac{5}{3}g_3^2-\frac{3}{4}g_2^2-\frac{1}{60}g_1^2\right)\right),\\
g_{U_{j,1}}=&\delta_{j3}\,\sqrt{\frac{6}{5}}\tilde g_1\left(1+\frac{1}{32\pi^2}\left(-\frac{4}{3}g_3^2-\frac{4}{15}g_1^2\right)\right),\\
g_{U_{j,3}}=&\delta_{j3}\,\sqrt{2}\tilde g_3\left(1+\frac{1}{32\pi^2}\left(\frac{5}{3}g_3^2-\frac{4}{15}g_1^2\right)\right),\\
g_{D_{j,1}}=&\delta_{j3}\,\sqrt{\frac{6}{5}}\tilde g_1\left(1+\frac{1}{32\pi^2}\left(-\frac{4}{3}g_3^2-\frac{1}{15}g_1^2\right)\right),\\
g_{D_{j,3}}=&\delta_{j3}\,\sqrt{2}\tilde g_3\left(1+\frac{1}{32\pi^2}\left(\frac{5}{3}g_3^2-\frac{1}{15}g_1^2\right)\right),\\
g_{L_{j,1}}=&\delta_{j3}\,\sqrt{\frac{6}{5}}\tilde g_1\left(1+\frac{1}{32\pi^2}\left(-\frac{3}{4}g_2^2-\frac{3}{20}g_1^2\right)\right),\\
g_{L_{j,2}}=&\delta_{j3}\,\sqrt{2}\tilde g_2\left(1+\frac{1}{32\pi^2}\left(\frac{5}{4}g_2^2-\frac{3}{20}g_1^2\right)\right),\\
g_{E_{j,1}}=&\delta_{j3}\,\sqrt{\frac{6}{5}}\tilde g_1\left(1+\frac{1}{32\pi^2}\left(-\frac{3}{5}g_1^2\right)\right),\end{align*}
\begin{align*}
\gamma_{1,H,H}=&\left(\frac{3 \tilde{g_1}^2}{5}+\tilde{g_2}^2\right) \cos^2(2\alpha)-\frac{75 \tilde{g_2}^4+30 \tilde{g_2}^2 \tilde{g_3}^2+9 \tilde{g_3}^4}{400 \pi ^2},\\
\gamma_{1,H,Q}=&-\frac{3 \left(75 \tilde{g_2}^4+\tilde{g_3}^4\right)}{400 \pi ^2}+\frac{3}{5} \tilde{g_1}^2 \cos(2\alpha)+6 \left(\tilde{y}_t^2 \cos^2\alpha+\tilde{y}_b^2 \sin^2\alpha\right),\\
\gamma_{1,H,U}=&\frac{3 \tilde{g_3}^4}{100 \pi ^2}-3 \tilde{y}_t^2 \cos^2\alpha+\frac{3}{5} \tilde{g_1}^2 \cos(2 \alpha),\\
\gamma_{1,H,D}=&-\frac{3 \tilde{g_3}^4}{200 \pi ^2}+\frac{3}{5} \tilde{g_1}^2 \cos(2 \alpha)+6 \tilde{y}_b^2 \sin^2\alpha,\\
\gamma_{1,H,L}=&\frac{75 \tilde{g_2}^4+9 \tilde{g_3}^4}{400 \pi ^2}+\frac{3}{5} \tilde{g_1}^2 \cos(2 \alpha)-2 \tilde{y}_\tau^2 \sin^2\alpha,\\
\gamma_{1,H,E}=&-\frac{9 \tilde{g_3}^4}{200 \pi ^2}+\frac{3}{5} \tilde{g_1}^2 \cos(2 \alpha)+2 \tilde{y}_\tau^2 \sin^2\alpha,\\
\gamma_{1,Q,Q}=&\frac{3 \tilde{g_1}^2}{5}+3 \tilde{g_3}^2-\frac{333 \tilde{g_1}^4+675 \tilde{g_2}^4+\tilde{g_3}^4+6 \tilde{g_1}^2 \left(135 \tilde{g_2}^2+\tilde{g_3}^2\right)}{400 \pi ^2},\\
\gamma_{1,Q,U}=&\frac{3 \tilde{g_1}^2}{5}+\frac{18 \tilde{g_1}^4+\tilde{g_3}^4}{100 \pi ^2}-3 \tilde{y}_t^2,\\
\gamma_{1,Q,D}=&\frac{3 \tilde{g_1}^2}{5}-\frac{72 \tilde{g_1}^4+\tilde{g_3}^4}{200 \pi ^2}+6 \tilde{y}_b^2,\\
\gamma_{1,Q,L}=&\frac{3 \tilde{g_1}^2}{5}+\frac{3 \left(75 \tilde{g_2}^4+\tilde{g_3}^4\right)}{400 \pi ^2},\\
\gamma_{1,Q,E}=&\frac{3 \tilde{g_1}^2}{5}-\frac{3 \tilde{g_3}^4}{200 \pi ^2},\\
\gamma_{1,U,U}=&\frac{3 \tilde{g_1}^2}{5}+\frac{3 \tilde{g_3}^2}{4}-\frac{117 \tilde{g_1}^4+96 \tilde{g_1}^2 \tilde{g_3}^2+64 \tilde{g_3}^4}{1600 \pi ^2},\\
\gamma_{1,U,D}=&\frac{3 \tilde{g_1}^2}{5}+\frac{9 \tilde{g_1}^4+2 \tilde{g_3}^4}{100 \pi ^2},\\
\gamma_{1,U,L}=&\gamma_{1,D,E}=\frac{3 \tilde{g_1}^2}{5}-\frac{3 \tilde{g_3}^4}{100 \pi ^2},\\
\gamma_{1,U,E}=&\frac{3 \tilde{g_1}^2}{5}+\frac{3 \tilde{g_3}^4}{50 \pi ^2},\\
\gamma_{1,D,D}=&\frac{3 \tilde{g_1}^2}{5}+3 \tilde{g_3}^2-\frac{117 \tilde{g_1}^4+24 \tilde{g_1}^2 \tilde{g_3}^2+4 \tilde{g_3}^4}{400 \pi ^2},\\
\gamma_{1,D,L}=&\frac{3 \tilde{g_1}^2}{5}+\frac{3 \tilde{g_3}^4}{200 \pi ^2},\\
\gamma_{1,D,E}=&\frac{3 \tilde{g_1}^2}{5}-\frac{3 \tilde{g_3}^4}{100 \pi ^2},\\
\gamma_{1,L,L}=&\frac{3 \tilde{g_1}^2}{5}+\tilde{g_2}^2-\frac{75 \tilde{g_2}^4+30 \tilde{g_2}^2 \tilde{g_3}^2+9 \tilde{g_3}^4}{400 \pi ^2},\\
\gamma_{1,L,E}=&\frac{3 \tilde{g_1}^2}{5}+\frac{9 \tilde{g_3}^4}{200 \pi ^2}-2 \tilde{y}_\tau^2,
\end{align*}
\begin{align*}
\gamma_{1,E,E}=&\frac{3 \tilde{g_1}^2}{5}-\frac{9 \tilde{g_3}^4}{100 \pi ^2},\\
\gamma_{2,H,Q}=&-\frac{\tilde{g_2}^2 \tilde{g_3}^2}{40 \pi ^2}-2 \tilde{y}_t^2 \cos^2\alpha+\tilde{g_2}^2 \cos(2 \alpha)+2 \tilde{y}_b^2 \sin^2\alpha,\\
\gamma_{2,H,L}=&\frac{3 \tilde{g_2}^2 \tilde{g_3}^2}{40 \pi ^2}+\tilde{g_2}^2 \cos(2 \alpha),\\
\gamma_{2,Q,Q}=&\tilde{g_2}^2+\tilde{g_3}^2+\frac{-9 \tilde{g_1}^4+30 \tilde{g_1}^2 \tilde{g_2}^2-2 \left(\frac{3 \tilde{g_1}^2}{5}+\tilde{g_2}^2\right) \tilde{g_3}^2}{240 \pi ^2},\\
\gamma_{2,Q,L}=&\tilde{g_2}^2+\frac{\tilde{g_2}^2 \tilde{g_3}^2}{40 \pi ^2},\\
\gamma_{3,Q,U}=&\tilde{g_3}^2+\frac{15 \tilde{g_1}^4+8 \tilde{g_1}^2 \tilde{g_3}^2}{400 \pi ^2}-2 \tilde{y}_t^2,\\
\gamma_{3,Q,D}=&\tilde{g_3}^2+\frac{15 \tilde{g_1}^4-4 \tilde{g_1}^2 \tilde{g_3}^2}{400 \pi ^2}-2 \tilde{y}_b^2,\\
\gamma_{3,U,D}=&\tilde{g_3}^2+\frac{-15 \tilde{g_1}^4+16 \tilde{g_1}^2 \tilde{g_3}^2}{400 \pi ^2},\\
\lambda'_E=&\tilde{y}_b \tilde{y}_\tau,\\
\lambda''_{E}=&0.
\end{align*}

Concerning dimensionful scalar couplings, the change of scheme only affects the fermion masses, such that
\begin{align*}
 M_k=&\tilde M_k\left(1+\frac{g_k^2}{16\pi^2}C(G_k)\right),\quad k=1,2,3,\\
\mu=&\tilde\mu\left(1+\frac{1}{16\pi^2}\left(\frac{3}{4}g_2^2+\frac{3}{20}g_1^2\right)\right).
\end{align*}
In the previous formulae, $C(G_k)$ denotes the adjoint Casimir of the group $k$, e.g. $C(G_2)=2$.

The rest of the parameters are equivalent in the two schemes (at least at one-loop order) but the parameterizations of the triple scalar couplings in the MSSM and the low energies theories is different. In the MSSM, these couplings receive contributions from both $\mu$ and the a-terms, while in the Effective SUSY scenarios they are parameterized simply by  $a_u$, $c_d$, $c_l$ in the Lagrangians of eqs.~\eqref{eq:LagrangianMES} and \eqref{eq:LagrangianNMES}. The matching is as follows
\begin{align*}
a_u=&\cos\alpha\tilde a_u+\sin\alpha \tilde c_u+\tilde\mu\sin\alpha\tilde y_t,\\
c_d=&\cos\alpha\tilde c_d+\sin\alpha\tilde a_d+\tilde\mu\cos\alpha\tilde y_b,\\
c_l=&\cos\alpha\tilde c_l+\sin\alpha\tilde a_l+\tilde\mu\cos\alpha\tilde y_\tau,
\end{align*}
where the MSSM a-terms are parameterized as
\begin{align}
\nonumber {\cal L}_{MSSM}\supset& -\tilde a_u Q \epsilon H_u U^c-\tilde a_d Q H_d D^c-\tilde a_l L H_d E^c-\tilde c_u Q \epsilon {H_d^\dagger} U^c -\tilde c_d Q H^\dagger_u U^c-\tilde c_l L H^\dagger_u E^c\\
&+{\rm c.c.}\label{eq:MSSMaterms}
\end{align}


\section{Low scale threshold corrections\label{app:lowscale}}

This appendix summarizes the low-energy one-loop thresholds considered in the calculations presented in the paper. First, the top and bottom Yukawas are  computed from the top and bottom pole masses by including order $\alpha^2$ QCD corrections \cite{Chetyrkin:1999qi}:
\begin{align}
\frac{m(q)}{m_{phys}}=1+\frac{\alpha_s}{\pi}\,z_{m_{phys}}^{(1)}(q)+\frac{\alpha^2_s}{\pi^2}\,z_{m_{phys}}^{(2)}(q)+O(\alpha_s^3),
\label{eq:QCDthr0}
\end{align}
\begin{align}
\nonumber z_{m_{phys}}^{(1)}=&-\frac{4}{3}-l_{q m},\\
\label{eq:QCDthr}z_{m_{phys}}^{(2)}=&\frac{16}{9}\left(-0.51056+\frac{21}{32}l_{q m}+\frac{9}{32}l_{q m}^2\right)+4\left(-3.33026-\frac{185}{96}l_{q m}-\frac{11}{32}l_{q m}^2\right)\\
\nonumber &+2\left(1.56205+\frac{13}{24}l_{q m}+\frac{1}{8}l_{q m}^2\right)+\frac{2}{3}\left(-0.15535+\frac{13}{24}l_{q m}+\frac{1}{8}l_{q m}^2\right),\\
\nonumber l_{\mu m}=&\log\frac{q^2}{m^2_{phys}},
\end{align}
where $m_t(q)=y_t(q) v, m_b(q)=y_b(q) v$ are the $\overline{\rm MS}$ scale-dependent tree-level masses. The values used for the quark pole masses are $m_t=172.9\pm0.6\pm0.9$, $m_b=4.78^{+0.20}_{-.07}$ \cite{Nakamura:2012}.

At the lower Effective SUSY threshold, one-loop corrections for the gauge couplings and Yukawas are considered. Denoting the Standard Model couplings as $g_i$ and the couplings in the minimal and nonminimal Effective SUSY scenarios as $\overline g_i$, $\hat g_i$ , respectively, the following matching equations apply for the gauge couplings, obtained from the results of ref.~\cite{Hall:1980kf}:
\begin{align}
\nonumber\overline g_{1}=&g_{1}-\frac{g_1^3}{(4\pi)^2}\left(\frac{2}{5}\log\frac{\overline\mu}{q}+\frac{1}{10}\left(\frac{1}{3}\log\frac{\overline m_Q}{q}+\frac{8}{3}\log\frac{\overline m_U}{q}\right)\right),\\
\label{eq:MESgthr}\overline g_{2}=&g_{2}-\frac{g_2^3}{(4\pi)^2}\left(\frac{4}{3}\log\frac{\overline M_2}{q}+\frac{2}{3}\log\frac{\overline\mu}{q}+\frac{1}{2}\log\frac{\overline m_Q}{q}\right),\\
\nonumber\overline g_{3}=&g_{3}-\frac{g_3^3}{(4\pi)^2}\left(2\log\frac{\overline M_3}{q}+\frac{1}{6}\left(2\log\frac{\overline m_Q}{q}+\log\frac{\overline m_U}{q}\right)\right),
\end{align}
\begin{align}
\nonumber\hat g_{1}=&g_{1}-\frac{g_1^3}{(4\pi)^2}\left(\frac{2}{5}\log\frac{\hat\mu}{q}+\frac{1}{10}\left(\frac{1}{3}\log\frac{\hat m_Q}{q}+\frac{8}{3}\log\frac{\hat m_U}{q}+\frac{2}{3}\log\frac{\hat m_D}{q}+\log\frac{\hat m_L}{q}+2\log\frac{\hat m_E}{q}\right)\right),\\
\label{eq:nMESgthr}\hat g_{2}=&g_{2}-\frac{g_2^3}{(4\pi)^2}\left(\frac{4}{3}\log\frac{\hat M_2}{q}+\frac{2}{3}\log\frac{\hat \mu}{q}+\frac{1}{2}\log\frac{\hat m_Q}{q}+\frac{1}{6}\log\frac{\hat m_L}{q}\right),\\
\nonumber\hat g_{3}=&g_{3}-\frac{g_3^3}{(4\pi)^2}\left(2\log\frac{\hat M_3}{q}+\frac{1}{6}\left(2\log\frac{\hat m_Q}{q}+\log\frac{\hat m_U}{q}+\log\frac{\hat m_D}{q}\right)\right).
\end{align}
Concerning the top and bottom Yukawa couplings, threshold corrections can be obtained by adapting to the Effective SUSY scenarios the results of ref.~\cite{Pierce:1996zz} for the MSSM.  In the approximation that neglects contributions other than those coming from stops and gluinos, the following matching relations are applied to the Yukawa couplings at the Standard Model-Effective SUSY threshold --again, we use tildes for the minimal Effective SUSY scenario, hats for the nonminimal one, and normal notation for the Standard Model:
\begin{align}
\nonumber \overline y_t=&y_t+\frac{m_t}{v}\frac{g_3^2}{12\pi^2}\left\{B_1(\overline m_{\tilde g},\overline m_{\tilde t_1})+f_1(\overline m_{\tilde g},\overline m_{\tilde t_2})-\sin2\overline\theta_t\frac{\overline m_{\tilde g}}{m_t}(f_0(\overline m_{\tilde g},\overline m_{\tilde t_1})-f_0(\overline m_{\tilde g},\overline m_{\tilde t_2}))\right\},\\
\label{eq:MESythr} \overline y_b=&y_b+\frac{m_b}{v}\frac{g_3^2}{12\pi^2}f_1(\overline m_{\tilde g},\overline m_{\tilde b}),
\end{align}
\begin{align}
 \nonumber\hat y_t=&y_t+\frac{m_t}{v}\frac{g_3^2}{12\pi^2}\left\{f_1(\hat m_{\tilde g},\hat m_{\tilde t_1})+f_1(\hat m_{\tilde g},\hat m_{\tilde t_2})-\sin2\hat\theta_t\frac{\hat m_{\tilde g}}{m_t}(f_0(\hat m_{\tilde g},\hat m_{\tilde t_1})-f_0(\hat m_{\tilde g},\hat m_{\tilde t_2}))\right\},\\
\label{eq:nMESythr}\hat y_b=&y_b+\frac{m_b}{v}\frac{g_3^2}{12\pi^2}\left\{f_1(\hat m_{\tilde g},\hat m_{\tilde b_1})+f_1(\hat m_{\tilde g},\hat m_{\tilde b_2})-\sin2\hat\theta_b\frac{\hat m_{\tilde g}}{m_b}(f_0(\hat m_{\tilde g},\hat m_{\tilde b_1})-f_0(\hat m_{\tilde g},\hat m_{\tilde b_2}))\right\},
\end{align}
In the previous formulae,  $\overline m_{\tilde g},\hat m_{\tilde g} $ denote gluino masses; $\overline m_{\tilde t_1}, \overline  m_{\tilde t_2}, \overline  m_{\tilde b}$ are the two stop mass eigenvalues ($\overline m_{\tilde t_1}$ being the heaviest) and the sbottom mass in the minimal Effective SUSY scenario, and $\hat m_{\tilde t_i},\hat m_{\tilde b_i}$  stand for the top/sbottom masses in the nonminimal case. The formulae for the masses are provided in \S \ref{app:VCW}. $\overline \theta_t$, $\hat \theta_t$ and $\hat \theta_b$ are the stop and sbottom mixing angles in the different scenarios, given by
\begin{align*}
 &\tan 2\overline\theta_t=\frac{24 {\overline a_u} v}{12 ({\overline m^2_Q}-{\overline m^2_U})+v^2 \left(\overline \gamma _{1,H,Q}+4 \overline \gamma _{1,H,U}-3 \overline \gamma _{2,H,Q}\right)},
\end{align*}
and similarly for  $\tan2 \hat\theta_t$ changing bars with hats, and
\begin{align*}
 &\tan2\hat \theta_b=-\frac{24 {\hat c_d} v}{12 ({\hat m^2_D}-{\hat m^2_Q})+v^2 \left(2 \hat \gamma _{1,H,D}-\hat \gamma _{1,H,Q}-3 \hat \gamma _{2,H,Q}\right)}.
\end{align*}
The functions $f_0$ and $f_1$ are $f_0(m_1,m_2)=B_0(0,m_1,m_2),\,f_1(m_1,m_2)=B_1(0,m_1,m_2)$, with the functions $B_0$ and $B_1$ given in eq.~\eqref{eq:Bfunctions}

In \S ~\ref{sec:Higgs}, Higgs masses are plotted against physical stop/sbottom masses. The Higgs masses are obtained from the effective potential, whose computation is summarized in \S~\ref{app:VCW}, while the third generation squark masses are calculated by taking into account one-loop self-energy corrections. These are obtained by using again the MSSM results of ref.~\cite{Pierce:1996zz} in the approximation that neglects $g_1, g_2$, the Yukawas of the first two generations, light quark masses and the mixing of charginos and neutralinos; to adapt the results for Effective SUSY scenarios, the terms involving the heavy Higgs states in the MSSM are dropped and the MSSM Yukawas are appropriately matched with the Effective SUSY ones. This gives the following expressions for the physical stop mass-matrix ${\cal M}_{\tilde t}$, valid for both minimal and nonminimal Effective SUSY scenarios (so that one can omit bars and hats).
\begin{align}\label{eq:mstop1}
 {\cal M}^2_{\tilde t}&={\cal M}^2_{\tilde t}+\left[
\begin{array}{cc}
 \Delta M^2_{LL} & \Delta M^2_{LR}\\
\Delta M^2_{LR} & \Delta M^2_{RR}
\end{array}
\right],\end{align}\begin{align*}
 \Delta M^2_{LL} &=\frac{g_3^2}{6\pi^2}\left\{2m^2_{\tilde t_2}\left[\cos^2\theta_t B_1(m_{\tilde t_1},m_{\tilde t_2},0)+\sin^2\theta_t B_1(m_{\tilde t_2},m_{\tilde t_2},0)\right]+A_0(m_{\tilde g})+A_0(m_t)\right.\\
&\left.-(m^2_{\tilde t_2}-m^2_{\tilde g}-m^2_t)B_0(0,m_{\tilde g},m_t)\right\}-\frac{1}{16\pi^2}\left[\tilde y_t^2\sin^2\theta_tA_0(m_{\tilde t_1})+\tilde y_b^2A_0(m_{\tilde b})-2(\tilde y_t^2\right.\\
&\left.+\tilde y_b^2)A_0(\mu)\right]\!-\!\frac{\tilde y^2_t}{32\pi^2}\left[\Lambda\left(\theta_t,\beta\!-\!\frac{\pi}{2}\right)B_0(0,m_{\tilde t_1},0)+\Lambda\left(\theta_t\!-\!\frac{\pi}{2},\beta\!-\!\frac{\pi}{2}\right)B_0(m_{\tilde t_2},m_{\tilde t_2},m_Z)\right]\end{align*}\begin{align}
\nonumber&-\frac{1}{16\pi^2}(\tilde y_t^2m_t^2\sin^2\beta+(\tilde y_b\mu\sin\beta+\tilde a_d\cos\beta)^2B_0(0,m_{\tilde b},0)),\\
\nonumber\Delta M^2_{LR}&=-\frac{g_3^2}{6\pi^2}\cos\theta_t\sin\theta_t\left[(m^2_{\tilde t_1}+m^2_{\tilde t_2}) B_0(m_{\tilde t_2},m_{\tilde t_1},0)+2m^2_{\tilde t_2}B_0(m_{\tilde t_2},m_{\tilde t_2},0)\right]\\
\nonumber&-\frac{g_3^2}{3\pi^2}m_tm_{\tilde g}B_0(0,m_t,m_{\tilde g})-\frac{3\tilde y^2_t}{16\pi^2}\cos\theta_t\sin\theta_t A_0(m_{\tilde t_1})\\
\nonumber&-\frac{\tilde y^2_t}{32\pi^2}\left[\Omega\left(\theta_t,\beta+\frac{\pi}{2}\right)B_0(0,m_{\tilde t_1},0)+\Omega\left(-\theta_t,\beta\!+\!\frac{\pi}{2}\right)B_0(m_{\tilde t_2},m_{\tilde t_2},m_Z)\right]\\
&\label{eq:mstop2}-\frac{1}{16\pi^2}(\tilde y_t m_t\sin\beta(\tilde y_t\mu\cos\beta+\tilde a_u\sin\beta)B_0(0,m_{\tilde b},0)),\\
\nonumber\Delta M^2_{RR} &=\frac{g_3^2}{6\pi^2}\left\{2m^2_{\tilde t_2}\left[\sin^2\theta_t B_1(m_{\tilde t_2},m_{\tilde t_1},0)+\cos^2\theta_t B_1(m_{\tilde t_2},m_{\tilde t_2},0)\right]+A_0(m_{\tilde g})+A_0(m_t)\right.\\
&\nonumber\left.-(m^2_{\tilde t_2}-m^2_{\tilde g}-m^2_t)B_0(0,m_{\tilde g},m_t)\right\}-\frac{\tilde y^2_t}{16\pi^2}\left[\cos^2\theta_tA_0(m_{\tilde t_1})+A_0(m_{\tilde b})-4A_0(\mu)\right]\\
&\nonumber-\frac{\tilde y^2_t}{32\pi^2}\left[\Lambda\left(\frac{\pi}{2}-\theta_t,\beta-\frac{\pi}{2}\right)B_0(0,m_{\tilde t_1},0)+\Lambda\left(-\theta_t,\beta\!-\!\frac{\pi}{2}\right)B_0(m_{\tilde t_2},m_{\tilde t_2},m_Z)\right]\\
&\nonumber-\frac{1}{16\pi^2}(\tilde y_t\mu\cos\beta+\tilde a_u\sin\beta)^2B_0(0,m_{\tilde b},0),
\end{align}
In the formulae above,  $m_{\tilde t_1}$ and $m_{\tilde t_2}$ are the heavy and light tree-level stop eigenvalues, respectively, and $m_{\tilde b}$ is the average of the sbottom eigenvalues --or the single sbottom mass in minimal scenarios. Also, $\tilde y_t=y_t/\cos\alpha,\,\,\tilde y_b=y_b/\sin\alpha,\,\,\tilde a_u=a_u/\cos\alpha-\mu\tilde y_t\tan\alpha,\,\,\tilde a_d=c_d/\sin\alpha-\mu\tilde y_b\cot\alpha$; it should be recalled that, according to eq.~\eqref{eq:alphabeta}, $\sin\alpha=\cos\beta$. The functions $A_0$, $B_0$, $B_1$, $\Lambda$ and $\Omega$ at a renormalization scale $q$ are given by 

\begin{align}
\nonumber A_0(m)=&m^2\left(1-\log\frac{m^2}{q^2}\right),\\
\nonumber  B_0(p,m_1,m_2)=&-\log\frac{p^2}{q^2}-f_B(x_+)-f_B(x_-),\end{align}\begin{align}
\label{eq:Bfunctions}x_\pm=&\frac{p^2-m_2^2+m_1^2\pm\sqrt{(p^2-m_2^2+m_1^2)^2-4p^2(m_1^2-i\epsilon)}}{2p^2},\\
\nonumber f_B(x)=&\log(1-x)-x\log(1-x^{-1})-1,\\
\nonumber B_1(p,m_1,m_2)=&\frac{1}{2p^2}(A_0(m_2)-A_0(m_1)+(p^2+m^2_1-m^2_2)B_0(p,m_1,m_2)).
\end{align}
\begin{align*}
\Lambda(\theta,\beta)=&\left(2m_t \cos\beta\cos\theta-\left(\mu\sin\beta-\frac{a_u}{\tilde y_t}\cos\beta\right)\sin\theta\right)^2+\left(\mu\sin\beta-\frac{a_u}{\tilde y_t}\cos\beta\right)^2\sin^2\theta,\\
\Omega(\theta,\beta)=&2m^2_t\cos^2\beta\sin2\theta_t-2m_t\cos\beta\left(\mu\sin\beta-\frac{a_u}{\tilde y_t}\cos\beta\right).
\end{align*}
 The formulae for the sbottom mass matrix in the nonminimal Effective SUSY case can be obtained from the above expressions by doing the substitutions $m_t\leftrightarrow m_b,\,\tilde y_t\leftrightarrow \tilde y_b,\,\beta\leftrightarrow\pi/2-\beta,\,a_u\leftrightarrow a_d$; the minimal case can be recovered from the $LL$ element by setting the mixing angle $\theta_b$ to $-\frac{\pi}{2}$.


\section{Effective potential\label{app:VCW}}

The Higgs mass is calculated from the effective potential in the $\overline{\rm MS}$ scheme. The full one-loop contribution is taken into account, as well as two-loop contributions coming from diagrams involving scalar fields. Following ref.~\cite{Martin:2001vx},
\begin{align*}
 V=&\frac{1}{16\pi^2}V^{(1)}+\frac{1}{(16\pi^2)^2}V_S^{(2)},
\end{align*}
where the one-loop contribution is given in terms of the tree-level mass-squared eigenvalues of fermions, real scalars and vector fields in a given background as
\begin{align*}
V^{(1)}=&\frac{1}{4}\sum_i\left(\log\frac{m^2_{S,i}}{q^2}-\frac{3}{2}\right)-\frac{1}{2}\sum_j\left(\log\frac{m^2_{F,j}}{q^2}-\frac{3}{2}\right)+\frac{3}{4}\sum_k\left(\log\frac{m^2_{V,k}}{q^2}-\frac{5}{6}\right),
\end{align*}
while the 2-loop contribution due to scalars can be expressed as
\begin{align}
\nonumber V_S^{(2)}=&V^{(2)}_{SSS}+V^{(2)}_{SS},\\
\nonumber V^{(2)}_{SSS}=&\frac{1}{12}({\lambda'}_{ijk})^2f_{SSS}(m^2_{S,i},m^2_{S,j},m^2_{S,k}),\\
\nonumber V^{(2)}_{SS}=&\frac{1}{8}({\lambda'}_{iijj})^2f_{SS}(m^2_{S,i},m^2_{S,j}),\end{align}\begin{align}
\label{eq:V2S} f_{SSS}(x,y,z)=&-\frac{1}{2}(x-y-z)\log\frac{y}{q^2}\log\frac{z}{q^2}-\frac{1}{2}(y-x-z)\log\frac{x}{q^2}\log\frac{z}{q^2}\\
\nonumber &-\frac{1}{2}(z-x-y)\log\frac{x}{q^2}\log\frac{y}{q^2}-2x\log\frac{x}{q^2}-2y\log\frac{y}{q^2}-2z\log\frac{z}{q^2}\\
\nonumber &+\frac{5}{2}(x+y+z)+\frac{1}{2}\xi(x,y,z),\\
\nonumber \xi(x,y,z)=&R\left\{2\log\left[\frac{z+x-y-R}{2z}\right]\log\left[\frac{z+y-x-R}{2z}\right]-\log\frac{x}{z}\log\frac{y}{z}\right.\\
\nonumber &\left.-2{\rm Li}_2\left[\frac{z+x-y-R}{2z}\right]-2{\rm Li}_2\left[\frac{z+y-x-R}{2z}\right]+\frac{\pi^2}{3}\right\},\\
\nonumber R=&(x^2+y^2+z^2-2xy-2xz-2yz)^{1/2}.
\end{align}
where the indices $i,j,k$ are summed over, and the couplings $\lambda'_{ijk},\lambda'_{ijkl}$ are the trilinear and quartic couplings of the Lagrangian written in terms of real scalars and evaluated on a given background, in the basis $\{S'\}$ on which the mass matrices in the background are diagonal.
\begin{align*}
 {\cal L}\supset -\frac{1}{6}\lambda'_{ijk}S'_iS'_jS'_k-\frac{1}{24}\lambda'_{ijkl}S'_iS'_jS'_kS'_l.
\end{align*}

The tree-level mass matrices for supersymmetric fermion and scalars in the normal interaction basis for the nonminimal Effective SUSY scenario are given next. 
Let $\psi_-=(\frac{1}{2}(\lambda_{2}^1+i\lambda_{2}^2),h_d^-), \psi_+=(\frac{1}{2}(\lambda_{2}^1-i\lambda_{2}^2),-h_u^+)$ designate the chargino fields, $\psi_0=(h_u^0,h_d^0,\lambda_{2}^3,\lambda_1)$  the neutralinos, $\Phi_U=(U_L,U_R)$ the stops --with $U_R=U^c$ in the notation of the tables in \S \ref{sec:Lagrangians}-- etc. Writing the mass terms in the Lagrangian as
\begin{align*}
 {\cal L}\supset& -\frac{1}{2}M_3\lambda_3\lambda_3-\psi_+ M_{F,+} \psi_--\psi_0 M_{F,0}\psi_0+{\rm c.c.}-\Phi^*_U M^2_{S,U}\Phi_U-\Phi^*_D M^2_{S,D}\Phi_D-\Phi^*_\tau M^2_{S,\tau}\Phi_\tau\\
&-\left(m^2_L-\frac{v^2}{4}\gamma_{1,H,L}-\frac{v^2}{4}\gamma_{2,H,L}\right)\Phi^*_{\tilde\nu} \Phi_{\tilde\nu},
\end{align*}
then one has
\begin{align*}
 M_{F,+}=\left[
\begin{array}{cc}
 M_2 & \frac{v}{2}g_{H^*_2}\\
\frac{v}{2}g_{H_2} & -\mu
\end{array}
\right],\quad 
M_{F,0}=\left[
\begin{array}{cccc}
0 & \mu & -\frac{v}{2}g_{H_2} & \frac{v}{2}g_{H_1}\\
\mu & 0 & \frac{v}{2}g_{H^*_2} & -\frac{v}{2}g_{H^*_1}\\
-\frac{v}{2}g_{H_2} & \frac{v}{2}g_{H^*_2} & M_2 & 0\\
\frac{v}{2}g_{H_1} & -\frac{v}{2}g_{H^*_1} & 0 & M_1
\end{array}\right],
\end{align*}
\begin{align}
 \nonumber M^2_{S,U}=&\left[
\begin{array}{cc}
 m^2_Q-\frac{v^2}{4}\gamma_{2,H,Q}+\frac{v^2}{12}\gamma_{1,H,Q} & a_u v\\
a_u v & m^2_U-\frac{v^2}{3}\gamma_{1,H,U}
\end{array}
\right],\\
\label{eq:Massessc}M^2_{S,D}=&\left[
\begin{array}{cc}
 m^2_Q+\frac{v^2}{4}\gamma_{2,H,Q}+\frac{v^2}{12}\gamma_{1,H,Q} & c_d v\\
c_d v & m^2_D+\frac{v^2}{6}\gamma_{1,H,D}
\end{array}
\right],\\
\nonumber M^2_{S,\tau}=&\left[
\begin{array}{cc}
 m^2_L-\frac{v^2}{4}\gamma_{1,H,L}+\frac{v^2}{4}\gamma_{2,H,L} & c_l v\\
c_l v & m^2_E+\frac{v^2}{2}\gamma_{1,H,E}
\end{array}
\right].
\end{align}
The mass matrices in the minimal Effective SUSY scenario are as above for the fermions and the stops, while in the case of the sbottom the mass matrix of eq.~\eqref{eq:Massessc} collapses into its (1,1) element.

The couplings  $\lambda'_{ijk},\lambda'_{ijkl}$ in eq.~\eqref{eq:V2S} can be obtained from the Lagrangians in \S~\ref{sec:Lagrangians} by expressing them in terms of real scalar fields and going to the basis on which the mass matrices are diagonal.

\bibliographystyle{h-physrev}
\bibliography{ESphenoref}

\end{document}